\documentclass[reprint,prb,aps,amsmath,amssymb,floatfix,superscriptaddress,longbibliography]{revtex4-1}
\usepackage{dcolumn}
\usepackage{bm}
\usepackage{graphicx}
\usepackage{amsthm}
\usepackage{color}
\usepackage{xcolor}
\usepackage{verbatim}
\usepackage{esint}
\usepackage[english]{babel}
\usepackage{amsfonts}
\usepackage{slashed}
\usepackage{latexsym}
\usepackage{bm}
\usepackage[colorlinks = true,
            linkcolor = red,
            urlcolor  = blue,
            citecolor = magenta,
            anchorcolor = red]{hyperref}
\usepackage{esint}
\usepackage{soul}
\usepackage{cancel}
\usepackage[normalem]{ulem}
\usepackage{empheq}
\usepackage{dsfont}
\usepackage{amsmath}
\usepackage{ulem}
\usepackage{epsfig}
\usepackage{enumerate}
\definecolor{sapphire}{rgb}{0.03, 0.03, 0.41}

\DeclareMathAlphabet\mathbfcal{OMS}{cmsy}{b}{n}
\def\be{\begin{equation}}
\def\ee{\end{equation}}

\begin{document}

\title{From non-equilibrium Green's functions to quantum master equations for the density matrix and out-of-time-order correlators:
 steady state and adiabatic dynamics}

\author{Bibek Bhandari}
\affiliation{NEST, Scuola Normale Superiore and Istituto Nanoscienze-CNR, I-56126 Pisa, Italy}
\affiliation{Department of Physics and Astronomy, University of Rochester, Rochester, NY 14627, USA
}
\author{Rosario Fazio}
\affiliation{ICTP, Strada Costiera 11, I-34151 Trieste, Italy}
\affiliation{Dipartimento di Fisica, Universit\`a di Napoli ``Federico II'', Monte S. Angelo, I-80126 Napoli, Italy}

\author{Fabio Taddei}
\affiliation{NEST, Istituto Nanoscienze-CNR and Scuola Normale Superiore, I-56126 Pisa, Italy}

\author{Liliana Arrachea}
\affiliation{International Center for Advanced Studies, Escuela de Ciencia y Tecnolog\'ia, Universidad Nacional de San Mart\'in and ICIFI,  Avenida 25 de Mayo y Francia, 1650 Buenos Aires, Argentina}



             
\begin{abstract}
We consider a finite quantum system under slow driving and weakly coupled to thermal reservoirs at different temperatures. We  present a systematic derivation of the quantum master equation
for the density matrix and the out-of-time-order correlators. We start 
from the microscopic Hamiltonian and we formulate  the equations ruling the dynamics of these quantities by recourse to the Schwinger-Keldysh non-equilibrium Green's function formalism, performing a perturbative expansion in the coupling between the system and the reservoirs. We focus on  the {\em adiabatic} dynamics, which corresponds to considering the linear response  in the ratio between the relaxation time due to the system--reservoir coupling  and  the time scale associated to the driving. We 
calculate the particle and energy fluxes. We illustrate the formalism in the case of a qutrit coupled to bosonic reservoirs and of a pair of interacting quantum dots attached to fermionic reservoirs, also discussing the relevance of coherent effects.
\end{abstract}

\maketitle

\section{Introduction}
The study of heat transport and heat--work conversion in few-level open quantum systems under the action of slow time-dependent protocols  is a subject of active investigation for some time now. 
Examples are qubits \cite{hanggi,avron,lmm,adiam1,adiam2,brandner,adiam3,adiam4,thingna,karini,abiuso,adiageo}, harmonic oscillators \cite{miche,adiaos,greenphon,marpaz,freipaz,riku,liu2020sharp,kalantar2020definitions},  and quantum dots \cite{janine1,calvo,ludovico,ludovico-capone,ribe,soth} under slow cyclic driving, as well as nanomechanical\cite{pistolesi2008self,brand,bode,thomas,marun,vib-cool,lucas1,lucas2} and nanomagnetic\cite{switch,magnet2} degrees of freedom in contact to bosonic or fermionic baths, possibly with a temperature bias. 

In the context of open systems the concept ``adiabatic dynamics" has been introduced to define  the evolution of slowly driven systems through time-dependent parameters \cite{thouless83,brouwer98,zhou99,moskalets02,moskalets04,adia-janine}. It applies to the non-equilibrium regime where the typical time scale of the dynamics of the frozen Hamiltonian  for the full setup, including the driven system along with  the contact to the reservoirs and the reservoirs themselves is much faster than the characteristic time for the changes of this Hamiltonian. This motivates a linear-response treatment with respect to the rate of change of the time-dependent parameters \cite{ludovico,adiageo}.
Similar ideas  are beyond the adiabatic perturbation theory in closed systems\cite{weinberg2017adiabatic,Bukov,kolu}.

A widely used framework to analyze the non-equilibrium dynamics of a few-level system  weakly-coupled to reservoirs  is that based  on {\em master equations}.
The standard approach is the Lindblad formulation~\cite{limdblad} which has been used to study the dynamics of different systems in the field of cold atoms, optics, quantum information and condensed matter~\cite{kosloff,nitzan,cuetara,felipe,adiam1,adiam2,adiam3,adiam4,hanggi,strass,hofer2017markovian,gonzalez2017testing,marino,jin,de2018reconciliation}. The main strategy of this formulation relies on
the equation of motion for the reduced density matrix of the quantum system, with the degrees of freedom of the reservoirs traced away. Another route to  derive  the master equation
is to calculate the dynamics of the mean values of the matrix elements of the density matrix by treating the coupling between the system and the reservoirs in 
perturbation theory within Schwinger-Keldysh contour. This implies considering a contour that evolves forwards and then
backwards with respect to an initial time $t_0 \rightarrow -\infty$.
The procedure was introduced in Ref.~\onlinecite{mak0} for a metallic island, in Refs.~\onlinecite{mak1,mak2} for a single-level quantum dot in the stationary regime, and extended to time-dependent scenarios in Refs.~\onlinecite{mak3,mak4,janine,janine2,dong}. A different formalism was also recently proposed~\cite{wej1}
and extended to time-dependent systems~\cite{wej2}. 

\begin{figure}
	\centering
	\includegraphics[width=\columnwidth]{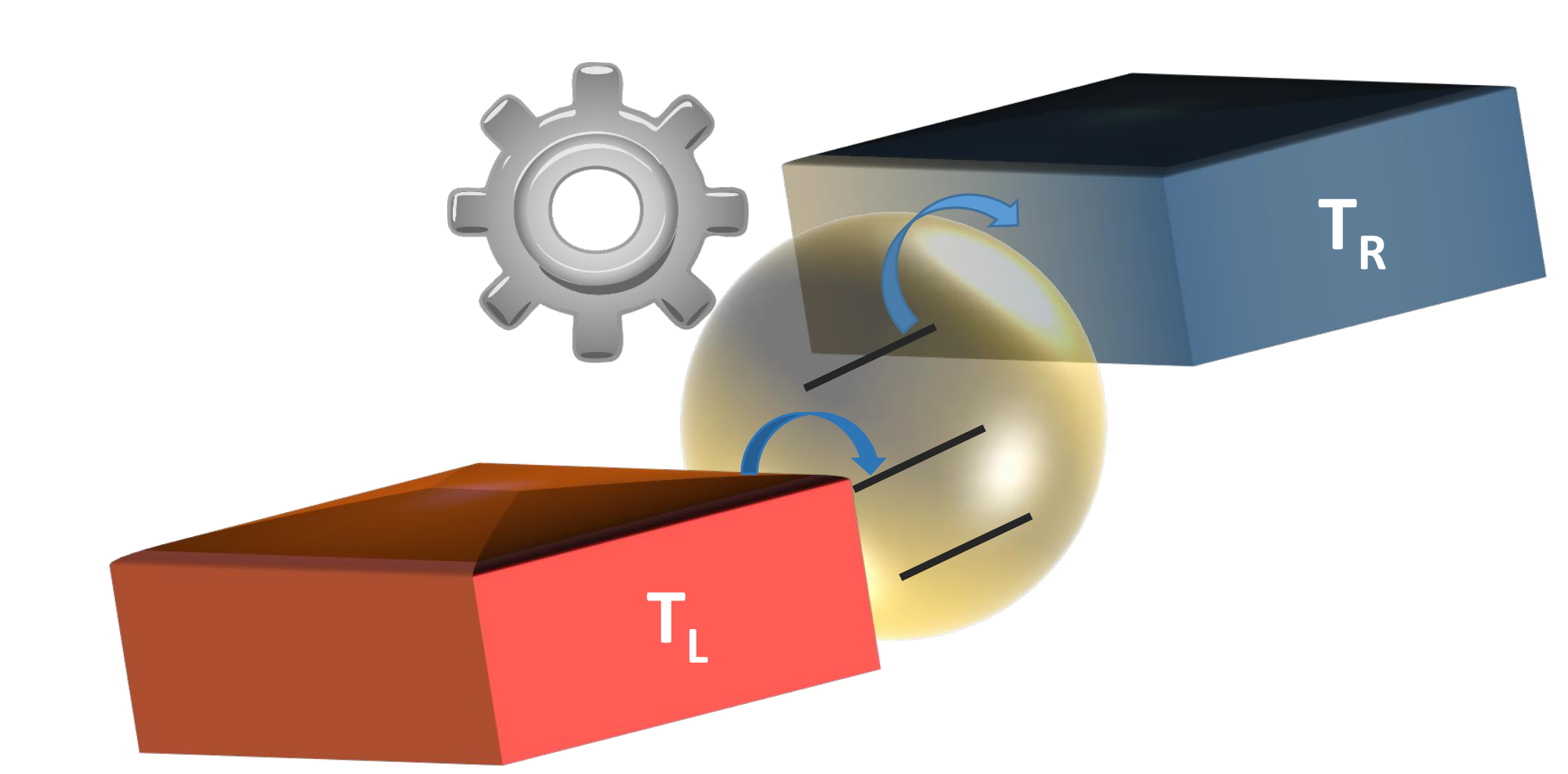}
	\caption{A N-level adiabatically driven system in contact with two reservoirs at different temperatures, $T_{\rm L}=T+\Delta T$ and $T_{\rm R}=T-\Delta T$.}
	\label{fig:sketch}
\end{figure}

Here, we present an alternative derivation of the master equations for the density matrix. We rely on the non-equilibrium Green's function  formalism combined with 
suitable analytical continuations\cite{jauho,rammer,greenelec1,stefa}. 
We focus on an adiabatically-driven $N$-level system weakly coupled to thermal reservoirs at different temperatures, see Fig.~\ref{fig:sketch}. We extend the procedure to the calculation of master equations for out-of-time-order correlators (OTOC), which
are currently under active investigation in the context of a variety of physical problems\cite{kitaev,maldacena,ioffe,sachdev,syzranov2018,gonza}.
OTOCs are considered  as good  witnesses  of  scrambling dynamics in many-body systems. In systems described by non-integrable  Hamiltonians,  OTOC's  are  expected to grow as a function of time 
\cite{kitaev,maldacena,ioffe,sachdev}. In systems coupled to thermal baths they stabilize after some time \cite{syzranov2018,gonza} and tend to an asymptotic value. The formalism we describe here enables  the analysis of these correlation functions in non-equilibrium situations, where the system is under slow driving and the reservoirs have a thermal or a chemical potential bias.

The paper is organized as follows. In the next section we will present the model for a periodically-driven quantum system. There, we will study the dynamics of the density matrix and the OTOCs. In section~\ref{secadia}, we will perform an adiabatic expansion to obtain full adiabatic master equations for density matrix as well as OTOCs. We will also derive the frozen and adiabatic contributions to the charge and energy currents in terms of the density matrix. In order to illustrate the general formulation, in section~\ref{exam} we put forward two different examples: in the first one we shall study a driven qutrit in contact with bosonic reservoirs, and in the second example we will consider a driven quantum dot system attached to fermionic reservoirs. Section \ref{conclu} is devoted to summary and conclusions. Some technical details are presented in appendices.

\section{General formalism}
\label{gf}
We present here the derivation of the master equation from the non-equilibrium Green's function formalism combined with the analytical continuation procedure known as Langreth theorem \cite{rammer,jauho}. 
\subsection{Model}
\label{sec:model}

We consider a driven quantum system S, which depends on time through a set of time-dependent parameters ${\bf X }(t)= \left(X_1(t),\ldots, X_N(t)\right)$, 
 described by the Hamiltonian 
\begin{equation}
{\cal H}_{\rm S}(t)\equiv {\cal H}_{\rm S}({\bf X}(t)).
\label{sysH1}
\end{equation}
In general, the system Hamiltonian contains one or more subsystems with multiple degrees of freedom, expressed in a convenient basis, which expands the $N$-dimensional Hilbert state. 
For example, 
in Section~\ref{exam} we consider a  qutrit (characterized by the three levels $|0\rangle$, $|1\rangle$ and $|2\rangle$) with time-dependent energies and time-dependent transitions between the different levels. We also consider  two 
coupled quantum dots of spinless fermions with time-dependent gate voltages and tunneling elements, as well as inter-dot Coulomb interaction. In this case, each quantum dot defines a subsystem, and the degrees of freedom of each quantum dot are determined by the charge.  
The corresponding states of the basis are four and read $|s \rangle \equiv | 0, 0\rangle,\; |1, 0 \rangle, \; |0,1 \rangle, \; |1,1 \rangle$.

The system S is coupled to a set of $N_r$ reservoirs described by the Hamiltonian
\begin{equation}
\mathcal{H}_{\rm B}=\sum_{\alpha=1}^{N_r} \sum_{k\alpha}\epsilon_{k\alpha}\hat{b}_{k\alpha}^\dagger\hat{b}_{k\alpha},
\label{Hb}
\end{equation}
where the operators $\hat{b}_{k\alpha}^\dagger$ and $\hat{b}_{k\alpha}$ (relative to an excitation in the bath $\alpha$ with momentum $k$) may satisfy bosonic or fermionic statistics. For the case of bosons, we focus on bosonic excitations, like phonons or photons. For the case of fermions, we focus on electron systems with a finite chemical potential. The contact between the driven system and the baths is given by the Hamiltonians
\begin{eqnarray}
\mathcal{H}_{\rm C}^{\rm (I)}&=&\sum_{s,s^{\prime}}\sum_{k,\alpha}V_{k\alpha} \hat{\pi}_{s,s^{\prime}}^{\alpha}  \left( \hat{b}_{k\alpha}^\dagger +\hat{b}_{k\alpha} \right), \nonumber \\
\mathcal{H}_{\rm C}^{\rm (II)}&=&\sum_{s,s^{\prime}}\sum_{k,\alpha} \left( V_{k\alpha} \hat{b}_{k\alpha}^\dagger \hat{\pi}_{s,s^{\prime}}^{\alpha}+H.c.\right),
\label{Hcon}
\end{eqnarray}
where $V_{k\alpha}$ is the coupling strength between the system and reservoir $\alpha$.
The structure of the Hamiltonian $\mathcal{H}_{\rm C}^{\rm (I)}$ corresponds to changing  $s$ in the central system while creating or destroying a quasiparticle in the bath and it
is a natural coupling in the case of reservoirs modeled by harmonic oscillators \cite{cal-leg}. Instead, $\mathcal{H}_{\rm C}^{\rm (II)}$ implies the creation of a particle (or quasiparticle) in the bath while changing $s$ of the central system. Usually, in the case of fermionic systems and reservoirs, such a term naturally describes a tunneling process where a fermion is destroyed in the system and created in the reservoir and vice-versa. Albeit, that type of coupling is also used in the case of $N$-level systems coupled to bosonic reservoirs.
In the derivation of the master equations we will consider, for the case of a bosonic bath, the Hamiltonian $\mathcal{H}_{\rm C}^{\rm (I)}$ and we will indicate how to get from them the corresponding equations for $\mathcal{H}_{\rm C}^{\rm (II)}$. For fermionic baths we will consider $\mathcal{H}_{\rm C}^{\rm (II)}$.
The operators $\hat{\pi}_{s,s^{\prime}}^{\alpha} = \eta^\alpha_{s,s'} |s\rangle\langle s'|$ are defined on the basis $|s \rangle$ associated to 
the degrees of freedom of the system and may be restricted by selection rules and by the Pauli principle in the case of fermionic systems. For instance, in the case of the two coupled quantum dots of spinless fermions 
that we will analyze in
Section~\ref{Sec_cqd}, where each quantum dot is connected to one fermionic reservoir through a tunnel coupling, 
these are $\hat{\pi}_{0,1}^{(1)} =\left[\hat{\pi}_{1,1}^{(1)}\right]^{\dagger} =\sum_{\ell=0,1}\eta^{(1)}_{0,1} |0, \ell \rangle\langle 1, \ell|$  for the quantum dot $(1)$ and 
$\hat{\pi}_{0,1}^{(2)} =\sum_{\ell=0,1}\left[\hat{\pi}_{1,1}^{(2)}\right]^{\dagger}= \eta^{(2)}_{0,1} |\ell, 0 \rangle\langle \ell, 1| $ for the quantum dot $(2)$.

The Hamiltonian for the system   at any time $t$ determining the value ${\bf X}$ of the time-dependent parameters can be diagonalized
by a unitary matrix $\hat{U}({\bf X})$ as follows
\begin{equation}
\tilde{\mathcal{H}}_{\rm S}({\bf X})=\hat{U}({\bf X}) {\cal H}_{\rm S}({\bf X})\hat{U}^\dagger({\bf X})=\sum_l \varepsilon_l({\bf X}) \hat{\rho}_{ll}({\bf X})
\label{eq:hamil_diagonal}
\end{equation}
where $\hat{\rho}_{lj}({\bf X})=|l({\bf X})\rangle\langle j({\bf X}) |$ is the density matrix expressed in the basis of the {\em instantaneous eigenstates} of the Hamiltonian, being 
\begin{equation}
\tilde{\mathcal{H}}_{\rm S}({\bf X})|l({\bf X})\rangle =\varepsilon_l({\bf X})|l({\bf X})\rangle.
\label{eq:vectorstate}
\end{equation}
We stress that this basis depends on time through the time-dependence of the parameters ${\bf X}$.
We define $\hat{\pi}_{\alpha} =\sum_{s,s^{\prime}} \hat{\pi}_{s,s^{\prime}}^{\alpha} $ and we express the contact
 Hamiltonian  in the instantaneous basis as follows
\begin{equation}
\label{eq:con_diag}
\tilde{\mathcal{H}}^{\rm (I,II)}_{\rm C}({\bf X})=\sum_{k,\alpha}\sum_{ l,j}V_{k\alpha}  \Big[\lambda_{\alpha,lj}({\bf X})\hat{b}_{k\alpha}^\dagger \hat{\rho}_{lj}+ \overline{\lambda}_{\alpha,lj}({\bf X})\hat{\rho}_{lj}\hat{b}_{k\alpha}\Big],
\end{equation}
where, for the case $\mathcal{H}_{\rm C}^{\rm (I)}$ we have
\begin{equation}
 \lambda_{\alpha,lj}({\bf X})=\overline{\lambda}_{\alpha,lj}({\bf X})
 =\left[ \hat{U}({\bf X})
\hat{\pi}_{\alpha}\hat{U}^\dagger({\bf X})\right]_{l,j},
\label{lambdab}
 \end{equation}
 while for the case $\mathcal{H}_{\rm C}^{\rm (II)}$ we have
   \begin{eqnarray}\label{lambdaf}
  \lambda_{\alpha,lj}({\bf X})&=&\left[\hat{U}({\bf X})\hat{\pi}_\alpha\hat{U}^\dagger({\bf X})\right]_{l,j},\nonumber \\
  \overline{\lambda}_{\alpha,lj}({\bf X})&=&\left[\hat{U}({\bf X})\hat{\pi}_\alpha^{\dagger} \hat{U}^\dagger({\bf X})\right]_{l,j}.
  \end{eqnarray}
\subsection{Dynamics of the density matrix and of the out-of-time-order correlator (OTOC)}\label{dynden}
The derivation of the equation of motion governing the long-time dynamics of the density matrix and of the out-of-time-order correlator, in the limit of weak coupling to the reservoirs, follows similar lines and we will treat the two cases in parallel.
``Long-time" refers to the regime beyond the transient associated to the switching-on of the coupling between system and reservoirs. 
 
We start by noticing that any observable ${\cal O}$, which depends on the degrees of freedom of the system, can be expressed as follows
\begin{equation} \label{o}
{\cal O}(t) = \sum_{l,j} O_{lj}(t) \hat{\rho}_{lj},
\end{equation}
where $O_{lj}(t)=\langle l | {\cal O}(t) |j \rangle $ are the matrix elements of the operator ${\cal O}(t)$ in the instantaneous eigenstates basis. The
expectation value  of this observable at a given time $t$ is 
\begin{equation} \label{ot}
\langle {\cal O} \rangle (t) = \mbox{Tr}\left[\hat{\rho}^{\rm tot} (t) {\cal O}(t) \right] = \sum_{lj} O_{lj}(t) \rho_{lj}(t),
\end{equation}
where we define the density matrix as
\begin{equation}\label{rho}
\rho_{lj} (t) = \mbox{Tr}\left[\hat{\rho}^{\rm tot} (t)  \hat{\rho}_{lj} \right].
\end{equation}
In Eqs.~(\ref{ot}) and (\ref{rho}) $\hat{\rho}^{\rm tot}(t)$ is the state of the full system coupled to the baths, which is described by the
Hamiltonian ${\cal H}(t)= \tilde{\cal H}_S(t)+ \mathcal{H}_{\rm B} + \tilde{\mathcal{H}}_{\rm C}(t)$. We see that the dynamics of $\langle {\cal O} \rangle (t)$ is determined by the 
evolution of the matrix elements of the operator in the basis of the instantaneous eigenstates of ${\cal H}_S(t)$ and the 
dynamics of 
the density matrix $\rho_{lj}(t)$. The latter depends on the full Hamiltonian ${\cal H}(t)$. 

Changing to the Heisenberg representation with respect to ${\cal H}$, $\hat{\rho}_{ij}^{\cal H}(t)= {\cal U}^{\dagger}(t,t_0)   \hat{\rho}_{ij} {\cal U}(t,t_0)$, the matrix elements of this operator are written as
\begin{equation}\label{rhoij0}
\rho_{lj}(t) = \mbox{Tr}\left[ \hat{\rho}_0\;  \hat{\rho}_{lj}^{\cal H}(t) \right],
\end{equation}
with $\hat{\rho}^{\rm tot} (t) = {\cal U}(t,t_0) \; \hat{\rho}_0 \; {\cal U}^{\dagger}(t,t_0)$, $\hat{\rho}_0$ being the state at the initial time $t_0$, where ${\cal U}(t,t_0)= \hat{T} \left\{ \exp^{-i/\hbar \int_{t_0}^t dt^{\prime} {\cal H} (t^{\prime}) } \right\}$ is the
evolution operator, being $\hat{T}$ the time-order operator. 
 
We define 
the OTOC between observables at time $t$, relative to a reference time $t_r$, as follows,
\begin{equation}
K(t)=\left\langle\mathcal{O}^{\cal H}_{A}(t)\mathcal{O}^{\cal H}_{B}(t_r)\mathcal{O}^{\cal H}_{C}(t)\mathcal{O}^{\cal H}_{D}(t_r)\right\rangle.
\label{eq:otoc_main}
\end{equation}
Here $\mathcal{O}_A,~\mathcal{O}_B,~\mathcal{O}_C$ and $\mathcal{O}_D$ are Hermitian operators depending on the degrees of freedom of the system S expressed in the Heisenberg picture
with respect to ${\cal H}$. 
We expand $\mathcal{O}_{A}(t)$ and $\mathcal{O}_{C}(t)$ 
as in Eq. (\ref{o}). The corresponding matrix elements are denoted, respectively, as $O_{A,lj}(t)$ and $O_{C,lj}(t)$. In this representation, the 
 OTOC of Eq.~(\ref{eq:otoc_main}) reads\cite{syzranov2018}
\begin{eqnarray}
K(t) &=& \sum_{lj l^{\prime} j^{\prime}}O_{A,lj}(t)O_{C,l^{\prime} j^{\prime}}(t) K_{lj, l^{\prime} j^{\prime}}(t), 
\end{eqnarray}
where we have introduced the OTOC operator 
\begin{equation}
\label{kk}
\hat{K}_{lj, l^{\prime} j^{\prime}}^{\cal H}(t) = \hat{\rho}_{lj}^{\cal H}(t)\mathcal{O}^{\cal H}_B(t_r)\hat{\rho}_{l^{\prime} j^{\prime}}^{\cal H}(t)\mathcal{O}^{\cal H}_D(t_r)
\end{equation}
and its corresponding mean value $K_{lj,l^{\prime} j^{\prime}}(t)= \left\langle\hat{K}_{lj,l^{\prime} j^{\prime}}^{\cal H}(t) \right\rangle$.

We now introduce the definitions of mixed lesser Green's functions for time correlations between the bath and the  density/OTOC operators, which we denote with $G$/$\mathcal{G}$
\begin{eqnarray}\label{gmix}
G_{lj,k\alpha}^{<} (t,t')&=&\pm i\left\langle b_{k\alpha}^{\dagger {\cal H}}(t')\hat{\rho}^{\cal H}_{lj}(t)\right\rangle, \nonumber \\
G_{k\alpha,lj}^{<}(t,t')&=&\pm i\left\langle \hat{\rho}^{\cal H}_{jl}(t') b^{\cal H}_{k\alpha} (t)\right\rangle, \nonumber \\
{\cal G}^{ <}_{k\alpha;ljl^{\prime} j^{\prime}}(t,t')&=&\pm i \left\langle \hat{T}_K\left[\hat{K}^{\cal H}_{lj,l^{\prime} j^{\prime}}(t') \right]^{\dagger}\hat{b}_{k\alpha}^{\cal H}(t)\right\rangle,\nonumber \\
{\cal G}^{ <}_{ljl^{\prime} j^{\prime};k\alpha}(t,t')&=&\pm i \left\langle \hat{T}_K\hat{b}_{k\alpha}^{\dagger{\cal H}}(t')\hat{K}^{\cal H}_{lj,l^{\prime} j^{\prime}}(t)\right\rangle.
\end{eqnarray}
For fermionic systems,
the upper/lower sign applies to many-body states such that $|l\rangle,\; |j\rangle$, as well as $|l^{\prime}\rangle,\; |j^{\prime}\rangle$, differ in odd/even number of particles. For bosonic systems, it corresponds the lower sign.
The operator $\hat{T}_K$ denotes  time-ordering along Schwinger-Keldysh contour, which starts at $-\infty$, evolves forwards  towards $+\infty$ and the backwards to $-\infty$\cite{jauho,rammer}. In the expressions for the OTOC, there are four operators at the times $t$ and  $t_r$. Hence, this contour in extended in order to include two of these contours\cite{ioffe}, as explained in Appendix~\ref{app:mean_otoc}.


Calculating the evolution of $ \hat{\rho}_{lj}^{\cal H}(t)$, taking the mean value with respect to $\hat{\rho}_0$ as in Eq. (\ref{rhoij0}) and introducing the definitions of the lesser Green's functions given in Eqs.  (\ref{gmix}), we get 
\begin{multline}
\frac{d \left\langle\hat{\rho}_{lj}\right\rangle}{dt}= \frac{i}{\hbar} \left[\varepsilon_{l}(t)- \varepsilon_j(t) \right] \left\langle\hat{\rho}_{lj}\right\rangle \pm \frac{1}{\hbar} \sum_{k,\alpha}V_{k\alpha}\times\\
\Big[\sum_m \lambda_{\alpha,ml}(t) G_{mj,k\alpha}^<(t,t)-\sum_n \lambda_{\alpha,jn}(t) G_{ln,k\alpha}^<(t,t)
\\
+\sum_m \overline{\lambda}_{\alpha,ml}(t) G_{k\alpha,jm}^<(t,t)-\sum_{n}\overline{\lambda}_{\alpha,jn}(t) G_{k\alpha,nl}^{<}(t,t)\Big],
\label{master1}
\end{multline}
where $\pm$ corresponds to fermionic and bosonic reservoirs, respectively, the first term in the right-hand side stems from $ \frac{i}{\hbar} \left\langle[\tilde{\mathcal{H}}^{\cal H}_{\rm S},\hat{\rho}^{\cal H}_{lj}]\right\rangle$ and we recall that $\varepsilon_l(t)\equiv \varepsilon({\bf X}(t))$.

Similarly, the equation of motion for the OTOC reads\cite{syzranov2018}
\begin{widetext}
\begin{multline}\label{master1otoc}
\frac{\partial \left\langle K_{ljl^{\prime} j^{\prime}}(t)\right\rangle}{\partial t}=\frac{i}{\hbar}\Big(\varepsilon_{l^{\prime}}(t)-\varepsilon_{j^{\prime}}(t)+\varepsilon_l(t)-\varepsilon_j(t)\Big) \left\langle K_{ljl^{\prime} j^{\prime}}(t)\right\rangle \pm \frac{1}{\hbar}\sum_{k\alpha}V_{k\alpha}\Big[\sum_m\lambda_{\alpha,ml}(t)\mathcal{G}^<_{mjl^{\prime} j^{\prime};k\alpha}(t,t)\\
-\sum_n\lambda_{\alpha,jn}(t)\mathcal{G}^{ <}_{lnl^{\prime} j^{\prime};k\alpha}(t,t)
+\sum_m \lambda_{\alpha,ml^{\prime}}(t)\mathcal{G}^{<}_{ljmj^{\prime};k\alpha}(t,t)-\sum_{n}\lambda_{\alpha,j^{\prime} n}(t)\mathcal{G}^{<}_{lj l^{\prime} n;k\alpha}(t,t)+\sum_m\bar{\lambda}_{\alpha,ml}(t)\mathcal{G}^{<}_{k\alpha;jm j^{\prime} l^{\prime} }(t,t)\\
-\sum_n\bar{\lambda}_{\alpha,jn}(t)\mathcal{G}^{<}_{k\alpha;nl j^{\prime} l^{\prime} }(t,t)+\sum_m \bar{\lambda}_{\alpha,ml^{\prime} }(t)\mathcal{G}^{<}_{k\alpha;jlj^{\prime} m}(t,t)-\sum_{n}\bar{\lambda}_{\alpha,j^{\prime} n}(t)\mathcal{G}^{<}_{k\alpha;jlnl^{\prime} }(t,t)\Big].
\end{multline}
 \end{widetext}

We now proceed with the line of argument presented in Refs.~\onlinecite{mak0,mak1,mak2}  to derive the master equations from Eqs.~(\ref{master1}) and (\ref{master1otoc}) based on the expansion of the coupling term
$V_{k \alpha}$. 
In our case  we  find it convenient to define non-equilibrium Green's functions for the  operators $\hat{\rho}_{l,j}$ and $\hat{K}_{lj,l^{\prime} j^{\prime}}$, in addition to the ones for the reservoirs and we proceed with the  derivation of  the Dyson equation at the lowest order in the couplings  in combination with
 Langreth theorem. These steps are similar to those followed in the study of quantum transport for strong coupling between system and reservoirs ~\cite{meir2,jauho}. To this end 
we introduce
the interaction representation with respect to the uncoupled Hamiltonian $h= \tilde{\cal H}_{\rm S}(t)+ \mathcal{H}_{\rm B} $. Therefore
\begin{equation}
 \hat{\rho}^{\cal H}_{ij}(t) = \hat{T}_K \left[ \exp\left\{-i/\hbar \int_{K} dt^{\prime} \tilde{\mathcal{H}}_{\rm C}^{h}(t^{\prime})\right\}\hat{\rho}_{ij}^{h}(t)\right],
\end{equation}
where the superscript $h$ denotes the interaction representation with respect to $h$ and we recall that $\hat{T}_K$ denotes time-ordering along the Schwinger-Keldysh contour. Furthermore, we set at the initial time $t_0=-\infty$,  $\rho_0 = \rho_{\rm S} \otimes \rho_{\rm B}$, where $\rho_{\rm S}, \; \rho_{\rm B}$ are the density operators of the uncoupled system and reservoirs, respectively.

The next step is  to evaluate the  Green's functions in Eq.~(\ref{master1}), up to the first order of perturbation theory in $V_{k\alpha}$.
It is convenient to introduce the definitions
\begin{align}\label{lambda0}
\Lambda_{mj}^{\alpha (0)}(t) &=\pm \sum_{k\alpha} V_{k\alpha} G_{mj,k\alpha}^<(t,t') ,\nonumber \\
\overline{\Lambda}_{mj}^{\alpha (0)}(t) &= \pm \sum_{k\alpha} V_{k\alpha} G_{k\alpha,mj}^<(t,t').
\end{align}
Using ``Langreth rule'',~\cite{jauho,rammer} we obtain the following expressions
\begin{multline}
\Lambda_{mj}^{\alpha (\kappa)}(t)  \simeq  \pm \sum_{u,v}\overline{\lambda}_{\alpha,uv}(t)\int_{-\infty}^\infty dt_1\Big(g_{mj,vu}^r(t,t_1)\\
\Sigma^{< (\kappa)}_{\alpha }(t_1,t)
+ g_{mj,vu}^{<}(t,t_1)\Sigma^{a (\kappa)}_{\alpha }(t_1,t)\Big),
\label{eq:lam}
\end{multline}
\begin{multline}
\overline{\Lambda}_{mj}^{\alpha (\kappa)}(t) 
\simeq  \pm
\sum_{u,v}\lambda_{\alpha,uv}(t)  \int_{-\infty}^\infty dt_1\Big(\Sigma^{r (\kappa)}_{\alpha}(t,t_1) \\
g_{uv,mj}^<(t_1,t)
+  \Sigma^{< (\kappa)}_{\alpha}(t,t_1) g_{uv,mj}^a(t_1,t)\Big),
\label{eq:overlam}
\end{multline}
where we have extended the definition of Eq. (\ref{lambda0}), corresponding to $\kappa=0$, to $\kappa=1$, by introducing the self-energies
\begin{equation}
\Sigma^{r,a,<(\kappa)}_{\alpha}(t,t^{\prime}) =  \int \frac{d\omega}{2\pi} e^{-i \omega (t-t^{\prime})} \Sigma^{r,a,< (\kappa)}_{\alpha}(\omega),
\end{equation}
which encode the coupling to the baths.
We can write the lesser self-energies as follows
\begin{eqnarray}
\Sigma^{< (\kappa)}_{\alpha}(\omega)= \pm i n_{\alpha}(\omega) \omega^{\kappa} \Gamma_{\alpha}(\omega),
 \end{eqnarray}
which depend on the spectral function
 \begin{equation}
  \Gamma_{\alpha}(\omega)= - 2 \mbox{Im}\left[\Sigma^{r (0)}_{\alpha}(\omega)\right] =2 \pi \sum_{k\alpha} \left|V_{k\alpha}\right|^2 
  \delta(\omega - \epsilon_{k\alpha}).
\end{equation}
Here $n_{\alpha}(\omega)$ denotes the Fermi-Dirac or Bose-Einstein distribution function for the case of fermionic or bosonic baths, respectively. 
 Importantly, the information on the temperature and chemical potential of a given reservoir $\alpha$ is only encoded in these functions.
 Notice that the index $\kappa$ in the previous expressions, denotes the different moments of the spectral function. In the previous expressions we have used the definitions of Eq. (\ref{lambdab}) and (\ref{lambdaf}) recalling that they depend on time through ${\bf X}(t)$.
 
In Eqs.~(\ref{eq:lam}) and (\ref{eq:overlam}), the lesser Green's functions are evaluated with respect to the uncoupled Hamiltonian $h$
\begin{eqnarray}\label{gsys}
g_{lj,vu}^<(t,t')&=&\pm i\left\langle \hat{\rho}^{h}_{uv} (t')\hat{\rho}^{h}_{lj}(t)\right\rangle,\nonumber \\
g_{lj,vu}^>(t,t')&=&- i\left\langle \hat{\rho}^{h}_{lj}(t) \hat{\rho}^{h}_{uv} (t')\right\rangle,
\end{eqnarray}
where $\hat{\rho}^{h}_{jl}(t) = [\hat{\rho}^{h}_{lj}(t)]^{\dagger}$, hence, $g^>_{\nu,\nu^{\prime}}(t,t^{\prime}) = \pm \left[g^<_{\nu^{\prime}, \nu}(t^{\prime},t) \right]^*$.
The corresponding retarded ones are
\begin{equation}
\label{gsys2}
g^r_{\nu,\nu^{\prime}}(t,t^{\prime}) = \theta(t-t^{\prime}) \left[ g^>_{\nu,\nu^{\prime}}(t,t^{\prime}) - g^<_{\nu,\nu^{\prime}}(t,t^{\prime}) \right],
\end{equation}
while the advanced Green's function is given by
$g^a_{\nu,\nu^{\prime}}(t,t^{\prime})=\left[g^r_{\nu^{\prime},\nu}(t^{\prime},t)\right]^*$.


In the case of the OTOC, we define
\begin{align}
 \Lambda^{\alpha, \rm OTOC}_{ljl^{\prime} j^{\prime}}(t) &= \pm \sum_{k\alpha} V_{k\alpha} {\cal G}_{ljl^{\prime} j^{\prime},k\alpha}^<(t,t),\nonumber \\
 \overline{\Lambda}^{\alpha, \rm OTOC}_{ljl^{\prime} j^{\prime}}(t) &= \pm \sum_{k\alpha} V_{k\alpha} {\cal G}_{k\alpha,ljl^{\prime} j^{\prime}}^<(t,t).
\end{align}
The evolution along Keldysh contour can be implemented by considering 
 an augmented contour \cite{ioffe}, which leads to a generalized Langreth rule, as explained in Appendix \ref{app:mean_otoc}. The counterparts to Eqs. (\ref{eq:lam}) and (\ref{eq:overlam}) for the OTOC functions read
\begin{multline}
\Lambda^{\alpha, \rm OTOC}_{ljl^{\prime} j^{\prime}} (t)\simeq \pm\int_{-\infty}^\infty dt_1  \sum_{u,v}\overline{\lambda}_{\alpha,uv}(t)\Big[g_{ljl^{\prime} j^{\prime},vu}^r(t,t_1)\\
\Sigma^{<(0)}_{\alpha}(t_1,t) +g_{ljl^{\prime} j^{\prime},vu}^<(t,t_1)\Sigma^{a(0)}(t_1,t)\Big],
\label{eq:mixed1}
\end{multline}
\begin{multline}
\overline{\Lambda}^{\alpha, \rm OTOC}_{ljl^{\prime} j^{\prime}}(t)  \simeq \pm\int_{-\infty}^\infty dt_1  \sum_{u,v} \lambda_{\alpha,uv}(t)\Big[\Sigma_{\alpha}^{r(0)}(t,t_1)\nonumber\\
g_{uv,ljl^{\prime} j^{\prime}}^<(t_1,t) +\Sigma_{\alpha}^{<(0)}(t,t_1)g_{uv,ljl^{\prime} j^{\prime}}^a(t_1,t)\Big],
\label{eq:mixed2}
\end{multline}
In order to obtain Eqs.~(\ref{eq:mixed1}), we fixed $t_r=t_0$, the initial time, and then we extended $t_0$ to $-\infty$ in the limits of the integrals. The retarded and lesser Green's functions for the OTOC are defined as
\begin{multline}
g^<_{ljl'j',vu}(t,t')\\
=\pm  i\left\langle \hat{T}_K\hat{\rho}_{uv}(t') \hat{\rho}_{lj}(t){\cal O}_B(t_r) \hat{\rho}_{l'j'}(t){\cal O}_D(t_r)\right\rangle\\
\pm i\left\langle \hat{T}_K\hat{\rho}_{lj}(t){\cal O}_B(t_r)\hat{\rho}_{uv}(t') \hat{\rho}_{l'j'}(t){\cal O}_D(t_r)\right\rangle,
\end{multline}
\begin{multline}
g^>_{ljl'j',vu}(t,t')\\
=- i\left\langle \hat{T}_K \hat{\rho}_{lj}(t)\hat{\rho}_{uv}(t'){\cal O}_B(t_r) \hat{\rho}_{l'j'}(t){\cal O}_D(t_r)\right\rangle\\
- i\left\langle \hat{T}_K\hat{\rho}_{lj}(t){\cal O}_B(t_r)\hat{\rho}_{l'j'}(t)\hat{\rho}_{uv}(t') {\cal O}_D(t_r)\right\rangle,
\end{multline}
The corresponding retarded  Green's function is
\begin{multline}
g^r_{ljl^\prime j^\prime,vu}(t,t')=\\
\theta(t-t')\bigg[g^>_{ljl^\prime j^\prime,vu}(t,t')-g^<_{ljl^\prime j^\prime,vu}(t,t') \bigg], 
\end{multline}
while the advanced one is given by $g_{\nu,\nu'}^a(t,t')=[ g_{\nu',\nu}^r(t',t)]^*$, in complete analogy with Eqs. (\ref{gsys})
\subsection{Dynamics of the particle and energy current between system and baths}
The time-resolved density matrix $\rho_{ij} (t)$ fully characterizes the dynamics of the local properties of the system.
We are also interested in evaluating the charge current $J_{\alpha}^{(c)}(t)$ (in the case of the fermionic reservoirs) as well as the energy current $J_{\alpha}^{(E)}(t)$ flowing between the system and the reservoirs. These quantities can also be calculated by recourse to Green's functions as follows

\begin{multline}
J_{\alpha}^{(c)}(t)=\frac{ie}{\hbar}\left\langle\Big[\mathcal{H},\mathcal{N}_\alpha\Big] \right\rangle
=\mp \frac{ e}{\hbar}\sum_{k}\sum_{m,n}V_{k\alpha} \times \\
\,\Big[\lambda_{\alpha,mn}(t)\,G_{mn,k\alpha}^<(t,t)
-\overline{\lambda}_{\alpha,mn}(t) \,G_{k\alpha,nm}^<(t,t)\Big],\nonumber
\end{multline}
\begin{multline}
J_{\alpha}^{(E)}(t)=\frac{i}{\hbar}\left\langle\Big[\mathcal{H},\mathcal{H}_\alpha\Big] \right\rangle
=\mp\frac{1}{\hbar}\sum_{k}\sum_{m,n}V_{k\alpha}\,\epsilon_{k\alpha}\times \\
\Big[\lambda_{\alpha,mn}(t)\,G_{mn,k\alpha}^<(t,t)
-\overline{\lambda}_{\alpha,mn}(t)\,G_{k\alpha,nm}^<(t,t)\Big],
\end{multline}
where the upper sign is for fermionic and lower sign for bosonic reservoirs.
Using  Eqs.~(\ref{eq:lam}) and (\ref{eq:overlam}), we can evaluate these currents to the lowest order in the coupling strength. 
The result is
\begin{align}\label{curr}
J_{\alpha}^{(c)}&=\frac{1}{\hbar}\bigg[\sum_{m,n}\overline{\lambda}_{\alpha,mn}(t)\overline{\Lambda}_{nm}^{\alpha (0)}(t)- \sum_{m,n}\lambda_{\alpha,mn}(t) \Lambda_{mn}^{\alpha (0)}(t)\bigg],
\nonumber \\
J_{\alpha}^{(E)}&=\frac{1}{\hbar}\bigg[\sum_{m,n}\overline{\lambda}_{\alpha,mn}(t) \overline{\Lambda}_{nm}^{\alpha (1)}(t)- \sum_{m,n}\lambda_{\alpha,mn}(t) \Lambda_{mn}^{\alpha (1) }(t)\bigg].
\end{align}
We see that the coefficients $\Lambda_{mn}^{\alpha (0)}(t)$ and $\overline{\Lambda}_{mn}^{\alpha (0)}(t)$ entering the equation of motion~(\ref{master1}) for the density matrix also enter the expression for the charge currents. 
Instead, the energy currents are determined by the coefficients  $\Lambda_{mn}^{\alpha (1)}(t)$ and $\overline{\Lambda}_{mn}^{\alpha (1)}(t)$ related to the first moment of the spectral function ($\kappa=1$).
Notice that the above expressions for the currents  are exact up to order $V_{k\alpha}^2$.


\section{Adiabatic dynamics}\label{secadia}
So far we have not introduced any assumptions regarding the nature of the time dependence. Here, we  focus on 
slow (adiabatic) driving, where the rate of change of the time-dependent parameters is small, which justifies treating the dynamics at different orders in these parameters.
More precisely, adiabatic driving is the regime where the typical time-scale $\tau$ associated to the driving is much larger than any other time-scale associated to the dynamics of the system coupled to the baths. 

\subsection{Green's function of the isolated system}\label{greeniso}
Here, we follow a treatment to evaluate the Green's functions of the isolated system along the line of Refs.~\onlinecite{ludovico,adiageo}, where linear response in the parameters $\dot{\bf X}$ was implemented.  
We recall that the  Green's function 
are evaluated with the operators expressed in the interaction picture with respect to $h=\tilde{\cal H}_{\rm S}(t)+ {\cal H}_{\rm B}$, which, for this particular function, is equivalent to the Heisenberg picture with 
respect to $\tilde{\cal H}_{\rm S}(t)$.

We consider the expansion of $\tilde{\cal H}_{\rm S}(t^{\prime})$ with respect to an ``observational time'' $t$,
\begin{eqnarray}\label{exp}
\tilde{{\cal H}}_{\rm S}(t')&=&\tilde{{\cal H}}^f_{\rm S}+ \delta \tilde{{\cal H}}_{\rm S}(t^{\prime}), \\ \nonumber
\delta \tilde{{\cal H}}_{\rm S}(t^{\prime}) &=& \sum_{n=1}^{\infty} \frac{(t^{\prime}- t)^n}{n ! } \frac{\partial \tilde{{\cal H}}_{\rm S}}{\partial {\bf X} } \cdot \frac{d^n {\bf X}}{dt^n}= \sum_{k=1}^N \xi_{k}(t^{\prime}) \hat{\rho}_{kk},
\end{eqnarray}
where $\tilde{{\cal H}}^f_{{\rm S}}$ is the Hamiltonian with the time  frozen at $t$ and
$\xi_{k}(t^{\prime})=-\sum_{n=1}^{\infty}\theta(\tau_{\rm ad} - |t-t^{\prime}|)(t^{\prime}-t)^n/n! d^n X_k/dt^n$, with $\tau_{\rm ad}<\tau$. 
In the latter expression, we 
introduce the function $\theta(\tau_{\rm ad} - |t-t^{\prime}|)$ to
indicate  that this expansion holds for time differences $|t^{\prime}-t|$, with respect to the observational time $t$, which are much smaller  than the typical time scale $\tau$ associated to the time-dependent parameters $\tau_{\rm ad}\ll\tau$.

We then change to 
the interaction representation with respect to $\tilde{{\cal H}}^f_{\rm S}$. We explain below the procedure followed for   the case of the  Green's function
\begin{eqnarray}\label{gless}
& & g_{ij,vu}^{<}(t_1,t_2)= \nonumber \\
 & & ~~~ - i \mbox{Tr} \left\{\hat{\rho}_0 \hat{T}_K \left[e^{-\frac{i}{\hbar}\int_K dt^{\prime}  \delta \tilde{{\cal H}}^f_{\rm S}(t^{\prime}) } \hat{\rho}^f_{lj}(t_{1}^+)  \hat{\rho}^f_{uv}(t_{2}^-) \right] \right\},
\end{eqnarray}
 where $t_{1}^+$ and $t_{2}^-$ indicates that the time $t_1$ is on the piece of the contour that starts in $-\infty$, while $t_2$ is on the piece of the contour that ends in $-\infty$.
 All the operators with the label $f$ are calculated in the Heisenberg representation of the frozen Hamiltonian $\tilde{{\cal H}}^f_{\rm S}$. In particular, 
\begin{equation} 
\hat{\rho}_{lj}^{f}(t^{\prime})=e^{\frac{i}{\hbar} t^{\prime} \tilde{{\cal H}}^f_{{\rm S}}}~\hat{\rho}_{lj}~e^{-\frac{i}{\hbar} t^{\prime} \tilde{{\cal H}}^f_{{\rm S}} }= e^{\frac{i}{\hbar} \epsilon_{lj} t^{\prime} }\hat{\rho}_{lj}
\label{eq:frogless}
\end{equation} 
with  $\epsilon^f_{j}$ being the eigenenergies of $\tilde{{\cal H}}^f_{{\rm S}}$
and $\epsilon_{lj}=\epsilon^f_{l}-\epsilon^f_{j}$. 
 Evaluating Eq. (\ref{gless}) up to linear order in the perturbation $\tilde{{\cal H}}^f_{\rm S}$ leads to
 \begin{eqnarray}\label{gless1}
& & g_{lj,vu}^{<}(t_1,t_2)\simeq  g_{lj,vu}^{<, f}(t_1,t_2) +\delta g_{lj,vu}^{<,f} (t_1,t_2),\nonumber \\
\end{eqnarray}
where the first term is the frozen component and reads
\begin{eqnarray}\label{frozen}
g_{lj,vu}^{<,f} (t_1,t_2) &= &\pm i  \delta_{lv} \left\langle\hat{\rho}_{uj}^f(t_1)\right\rangle e^{i \epsilon_{uv}(t_2-t_1)} \nonumber \\
&=& \pm i  \delta_{lv} \left\langle\hat{\rho}_{uj}^f(t_2)\right\rangle e^{i \epsilon_{jv}(t_2-t_1)},
\end{eqnarray}
while the second term is the correction up to linear order in $\delta \tilde{{\cal H}}^f_{\rm S}(t^{\prime})$ (see calculation in Appendix~\ref{app:den_ada}) and reads
\begin{eqnarray}\label{dg}
\delta g_{lj,vu}^{<,f}(t_1,t_2)
&= & -\frac{i}{\hbar}\,g_{lj,uv}^{<,f}(t_1,t_2) 
\bigg[
\int_{-\infty}^{t_1}dt' \xi_j(t')  \nonumber\\
 & &+  \int_{t_1}^{t_2}dt' \xi_v(t')-\int^{t_2}_{-\infty}dt' \xi_u(t')\bigg].
\end{eqnarray}
Notice that, in spite of the fact that some of the limits of the integrals  are defined to be $-\infty$, the functions $\xi_j (t')$ are different from zero only for $|t-t'|< \tau_{\rm ad}$.
Also notice that these functions, through Eqs.~(\ref{gless1}) and (\ref{dg}), enter the definitions of the functions of Eqs.~(\ref{eq:lam}) and (\ref{eq:overlam}) convoluted with the self-energy of the baths, which decay within the relaxation time due to the coupling to the bath, $\tau_{\rm rel}=\hbar/\Gamma_{\alpha}$.
 Hence, we identify $\tau_{\rm ad} \simeq \tau_{\rm rel}$.

Therefore, the validity of the present treatment in the description of the finite system coupled to the bath is restricted to $\tau_{\rm rel} < \tau$.
The {\em adiabatic} approximation consists in keeping the terms $\propto \dot{\bf X}$ in 
$\delta g_{lj,vu}^{<,f} (t_1,t_2)$
 under the assumption that the changes in ${\bf X}(t)$
take place within a time scale that is much larger than the typical time scale of the dynamics of the frozen system, hence $\tau_{\rm rel} \ll \tau$. 

A similar procedure can be followed to evaluate the Green's functions for the OTOC. In that case, the counterpart of Eq. (\ref{gless1}) is
\begin{equation}
g_{ljl'j',vu}^{<}(t_1,t_2) \simeq g_{ljl'j',vu}^{<, f}(t_1,t_2) +\delta g_{ljl'j',vu}^{<,f} (t_1,t_2),
\label{eq:lesstot_otoc}
\end{equation}
where the frozen term reads
\begin{multline}
g_{ljl'j',vu}^{<, f}(t_1,t_2)= 
 \pm i\, \Big[\delta_{lv} \left\langle K_{ujl'j'}^f(t_1)\right\rangle \\
 +\delta_{l'v} \left\langle K_{ljuj'}^f(t_1)\right\rangle\Big]e^{i\epsilon_{uv}(t_2-t_1)},
\label{eq:lessfr_otoc}
\end{multline}
while the linear order term in $\delta \tilde{{\cal H}}^f_{\rm S}(t^{\prime})$ is given by
\begin{multline}
\delta g_{ljl'j',vu}^{<,f}(t_1,t_2)=\pm \frac{1}{\hbar}\int_{-\infty}^\infty dt' \bigg(\left\langle \hat{K}_{ujl'j'}^f(t_1)\right\rangle \delta_{vl}\times\\
\Big[\theta(t_1-t')\xi_{j'j,l'v}(t')+\theta(t_2-t')\xi_{v,u}(t')\Big]+\left\langle \hat{K}_{ljuj'}^f(t_1)\right\rangle\\
\delta_{vl'}\Big[\theta(t_1-t')\xi_{j'j,lv}(t')+\theta(
t_2-t')\xi_{v,u}(t')\Big]\bigg)e^{i\epsilon_{vu}(t_1-t_2)} ,
\label{eq:lessad_otoc}
\end{multline}
where $\xi_{j'j,l'v}=\xi_{j'}+\xi_{j}-\xi_{l'}-\xi_{v}$ and $\xi_{v,u}=\xi_v-\xi_u$. 

The lesser functions in Eqs. (\ref{gless1}) and (\ref{eq:lesstot_otoc}) enter the master equations through Eqs.~(\ref{eq:lam}), (\ref{eq:overlam}), and (\ref{eq:mixed1}),  respectively. 
Notice that while the frozen components of these functions lead to a result $\propto V_{\alpha}^2$ for the $\Lambda$-functions, the adiabatic corrections $\delta g^<$ lead to a higher order correction
$\propto V_{\alpha}^2 \dot{\bf X}$. As we will further discuss below, this term can be neglected in comparison to others.

\subsection{Master equations}
\label{fame}
\subsubsection{Density matrix}
Our aim is to calculate the matrix elements of the density matrix up to linear order in $\dot{\bf X}$. Hence, we split them as follows,
\begin{equation}\label{p}
 \rho_{uj}(t)\equiv \rho_{uj}^f(t)+\rho_{uj}^a(t).
\end{equation}
In the previous equations, 
$\rho_{uj}^f$ is the solution of the frozen master equation for the density matrix, while 
 $\rho_{uj}^a(t)$ is the corresponding correction $\propto \dot{\bf X}$.
 
 The diagonal and off-diagonal terms of $\rho_{uj}$ are named, respectively, {\em populations} and {\em coherences} and are generally coupled (in what follow we will use the shorthand notation $p_u=\rho_{uu}$ for the populations). 
 By substituting Eq.~(\ref{gless}) into Eqs.~(\ref{eq:lam}) and (\ref{eq:overlam}), with Eq.~(\ref{frozen}) and the adiabatic approximation of Eq.~(\ref{dg}), the master equation that includes both frozen and adiabatic contributions can be written as
\\
\begin{multline}
  \frac{d \rho_{lj}}{dt}=\frac{i}{\hbar} \Big[\varepsilon_{l}(t)- \varepsilon_j(t) \Big] \rho_{lj}   +
  \sum_{mu,\alpha}\Big[ W_{ml,\alpha}^{ju}(t){\rho}_{mu}(t)\\
  +\tilde{W}_{jm,\alpha}^{ul}(t){\rho}_{um}(t)-W_{jm,\alpha}^{mu}(t){\rho}_{lu}(t) -\tilde{W}_{ml,\alpha}^{um}(t){\rho}_{uj}(t)\Big],
  \label{mas_sim}
\end{multline}
where we introduced  the transition rates
\begin{eqnarray}
W_{ml,\alpha}^{ju}(t)= W_{ml,\alpha}^{ju (f)}(t)+ \delta W_{ml,\alpha}^{ju}(t).
\end{eqnarray}
In deriving Eq.~(\ref{mas_sim}), we have neglected the level renormalization effects.

The explicit expressions for the frozen rates originated in the contribution of  Eq.~(\ref{frozen}) are  
\begin{multline}\label{ratef1}
W^{ju, (f)}_{ml,\alpha}= \lambda_{\alpha,ml}(t) \overline{\lambda}_{\alpha,ju}(t) \gamma^f_\alpha(\epsilon_{ju})/2 \\
+\overline{\lambda}_{\alpha,ml}(t)
\lambda_{\alpha,ju}(t) \tilde{\gamma}^f_\alpha(\epsilon_{uj})/2 ,
\end{multline}
\begin{multline}\label{ratef2}
\tilde{W}^{ul, (f)}_{jm,\alpha}=\lambda_{\alpha,jm}(t) \overline{\lambda}_{\alpha,ul}(t) \tilde{\gamma}^f_{\alpha}(\epsilon_{ul})/2\\
+\overline{\lambda}_{\alpha,jm}(t) \lambda_{\alpha,ul}(t) \gamma^f_{\alpha}(\epsilon_{lu})/2,
\end{multline}
where
\begin{eqnarray}\label{gammasf}
\gamma_\alpha^f(\epsilon)&=&\hbar^{-1}n_\alpha(\epsilon)\Gamma_\alpha(\epsilon), \nonumber \\
\tilde{\gamma}_\alpha^f(\epsilon)&=&\hbar^{-1}(1\mp n_\alpha(\epsilon))\Gamma_\alpha(\epsilon).
\end{eqnarray}
On the other hand, the adiabatic corrections to the transition rates 
$\delta W_{lj,\alpha}^{mu}(t)$,
which have their origin in Eq.~(\ref{dg}), can be evaluated in a similar manner (see Appendix~\ref{app:den_ada} for details). The latter are $\propto \Gamma_{\alpha} \dot{\bf X}$ within the adiabatic approximation. 

As mentioned before, in all the calculations leading to the master equations~(\ref{mas_sim}) we have considered the contact Hamiltonian in Eq.~(\ref{eq:con_diag}). Such master equations thus hold for both ${\cal H}_C^{\rm (I)}$ and ${\cal H}_C^{\rm (II)}$.
In the former case, the quantities $\lambda_{\alpha,lj}$ and $\overline{\lambda}_{\alpha,lj}$ entering the transition rates (\ref{ratef1}) and (\ref{ratef2}) are
defined by Eq.~(\ref{lambdab}), while in the latter they are defined by Eq.~(\ref{lambdaf}).

We now introduce the following schematic notation for Eq. (\ref{mas_sim}),
\begin{equation}
    \frac{d \mbox{\boldmath$\rho$}}{dt}= \frac{i}{\hbar} {\mbox{\boldmath$\epsilon$}}~{\mbox{\boldmath$\rho$}}+ {\bf W} ~{\mbox{\boldmath$\rho$}}.
\end{equation}
Splitting in this equation the density matrix elements and rates into their frozen and adiabatic components as in Eq.~(\ref{p}), we can make use of the fact that the frozen component satisfies
\begin{equation}
    0= \frac{i}{\hbar} {\mbox{\boldmath$\epsilon$}}~{\mbox{\boldmath$\rho$}}^f+ {\bf W}^f ~{\mbox{\boldmath$\rho$}}^f,
\label{mas_froz}
\end{equation}
to conclude that the following equation has to be fulfilled by the adiabatic components (keeping only linear-order terms in $\dot{\bf X}$),
\begin{equation}
     \frac{\partial{\mbox{\boldmath$\rho$}}^f}{\partial {\bf X}} \dot{\bf X}= \frac{i}{\hbar} {\mbox{\boldmath$\epsilon$}}~{\mbox{\boldmath$\rho$}}^a+ {\bf W}^f ~{\mbox{\boldmath$\rho$}}^a+ \mbox{\boldmath$\delta$}{\bf W}~ 
     {\mbox{\boldmath$\rho$}}^f.\label{mas_ad}
\end{equation}
These equations must be supplemented by the normalization of the populations $\sum_l p_{l}=1$.
Notice that the term in the left-hand side contains two components. One component is originated in the variation with respect to ${\bf X}$ of the matrix ${\bf M} = i {\mbox{\boldmath$\epsilon$}}/\hbar+{\bf W}^f$
entering Eq. (\ref{mas_froz}), 
while the other one is due to the change of the instantaneous eigenstates  as ${\bf X}$ changes [instantaneous eigenvalues and eigenstates are defined in Eq.~(\ref{eq:vectorstate})]. The contribution of these two terms in the  derivatives of the matrix elements of ${\mbox{\boldmath$\rho$}}^f$ with respect to ${\bf X}$ reads 
\begin{eqnarray}\label{change-adia}
\frac{\partial \rho_{l,j}^f}{\partial {\bf X}}& = & - \left[{\bf M}^{-1} \frac{\partial {\bf M}}{\partial{\bf X}} {\mbox{\boldmath$\rho$}}^f \right]_{l,j}\nonumber \\
& & 
+ \sum_{l^{\prime}} \left\{ A_{l^{\prime},l} \rho_{ l^{\prime}, j} -  A_{j,l^{\prime}} \rho_{ l, l^{\prime}} \right\},
\end{eqnarray}
being 
\begin{equation}
    A_{l^{\prime},l}= \frac{\langle l^{\prime}|\frac{\partial {\cal H}_{\rm S}}{\partial{\bf X}}|l\rangle}{\varepsilon_l-\varepsilon_{l^{\prime}}}, ~~~~l\neq l^{\prime}.
\end{equation}  
The second term of Eq. (\ref{change-adia}) is equivalent to the contribution of the gauge potential in the moving-frame introduced in the framework of the adiabatic perturbation theory for closed quantum systems\cite{weinberg2017adiabatic} (see details in Appendix \ref{apgau}). This term does not play any role when the master equation is reduced to a rate equation by taking into account the evolution of the populations only \cite{mak3,mak4,janine,adiageo}, but it has been considered in the adiabatic evolution of open quantum systems described by the Lindblad master equation\cite{adiam3,abiuso}.

On the other hand, as already mentioned in Section \ref{greeniso}, the contribution of the  terms collected in $\mbox{\boldmath$\delta$}{\bf W}~ {\mbox{\boldmath$\rho$}}^f$ in
the previous equation are effectively higher order in the parameters defining the perturbative treatment.
In fact, $\mbox{\boldmath$\delta$}{\bf W}$
is linear order in the rate amplitude $\Gamma_\alpha$ (which is in turn second order in the coupling $V_{k \alpha}$)  times linear order in the adiabatic expansion $\dot{\bf X}$.
Hence, we  neglect these terms in comparison to the ones containing ${\bf W}^f$, since these matrix elements are linear in $\Gamma_{\alpha}$ and $0$-th order in the adiabatic expansion. Therefore, keeping only the latter terms and neglecting the former ones, we get ${\mbox{\boldmath$\rho$}}^a \sim {\cal O}(\tau_{\rm rel}/\tau)$.
This reasoning is basically the same as that presented in Ref.~\onlinecite{mak3} and the result highlights the fact that the validity of the adiabatic treatment is restricted to variations of the driving in a time-scale much larger than the relaxation time of the system with the environment ($\tau \gg \tau_{\rm rel}$).

\subsubsection{OTOC}
Similarly, for the OTOC we introduce the decomposition
\begin{align}
K_{lj,l'j'}(t) \equiv K_{lj,l'j'}^f(t)+K_{lj,l'j'}^a(t),
 \label{eq:otoc_adfr}
\end{align}
where $K_{lj,l'j'}^f$ is the solution of the frozen master equation for the OTOC,  while 
 $K_{lj,l'j'}^a$ is the corresponding corrections $\propto \dot{\bf X}$.
The master equation for the OTOC can be derived in a similar manner as before, by substituting Eqs.~({\ref{eq:lesstot_otoc}}),~(\ref{eq:lessfr_otoc}) and~(\ref{eq:lessad_otoc}) into Eqs.~(\ref{eq:mixed1}) and (\ref{eq:mixed2}), obtaining
\begin{widetext}
\begin{multline}
\frac{dK_{lj,l'j'}(t)}{dt}=\frac{i}{\hbar} \Big[\varepsilon_{l^{\prime}}(t)-\varepsilon_{j^{\prime}}(t)+\varepsilon_l(t)-\varepsilon_j(t)\Big] K_{lj,l^{\prime} j^{\prime}}(t)
+ \sum_{mu, \alpha}\Big[{W}_{ml,\alpha}^{ju}(t){K}_{mu,l'j'}(t)
+{W}_{ml,\alpha}^{j'u}(t){K}_{mj,l'u}(t)
\\
+\tilde{W}_{jm,\alpha}^{ul}(t)K_{um,l'j'}(t)+\tilde{W}_{jm,\alpha}^{ul'}K_{lm,uj'}(t)+W_{ml',\alpha}^{ju}K_{lu,mj'}(t)+W_{ml',\alpha}^{j'u}K_{ljmu}(t)+\tilde{W}_{j'm,\alpha}^{ul}(t)K_{uj,l'm}(t)+\tilde{W}_{j'm,\alpha}^{ul'}(t)K_{lj,um}(t)\Big]\\
-\sum_{mu, \alpha}\Big[\tilde{W}_{ml,\alpha}^{um}(t){K}_{uj,l'j'}(t)+\tilde{W}_{ml,\alpha}^{ul^\prime}(t){K}_{mj,uj'}(t)
+{W}_{jm,\alpha}^{mu}K_{lu,l'j'}+W_{jm,\alpha}^{j'u}K_{lm,l'u}+\tilde{W}_{ml',\alpha}^{ul}(t)K_{uj,mj'}(t)\\
+\tilde{W}_{ml',\alpha}^{um}(t)K_{lj,uj'}(t)+{W}_{j'm,\alpha}^{ju}(t)K_{lu,l'm}(t)+{W}_{j'm,\alpha}^{mu}(t)K_{lj,l'u}(t)\Big]
,
\end{multline}
\end{widetext}
where 
the rates $W_{ml,\alpha}^{ju}(t)$ and $\tilde{W}_{ml,\alpha}^{ju}(t)$ are the same appearing in Eq.~(\ref{mas_sim})
(see the Appendix~\ref{app:tran_otoc} for details). Using a similar schematic notation as before, we have
\begin{equation}\label{master-otoc-fin}
   \frac{d{\bf K}}{dt}= \frac{i}{\hbar} {\mbox{\boldmath$\epsilon$}}^{\rm OTOC}~{\bf K}+ {\bf W}^{\rm OTOC} ~{\bf K}.  
\end{equation}
The procedure to formulate the adiabatic master equation for the OTOC is the same as the one for the density matrix. As before, the rates are  split 
into frozen and adiabatic components whose origin can be traced back to Eq.~(\ref{eq:lessfr_otoc}) and Eq.~(\ref{eq:lessad_otoc}) respectively,
\begin{equation}
{\bf W}^{\rm OTOC} = {\bf W}^{{\rm OTOC}, f}+\mbox{\boldmath$\delta$}{\bf W}^{\rm OTOC}.
\end{equation}
Introducing this decomposition, as well as the one in Eq.~(\ref{eq:otoc_adfr}) leads to the master equation for the steady state, describing  the long-time dynamics (for $t \rightarrow \infty$) of the frozen component,
\begin{equation}
   0= \frac{i}{\hbar} {\mbox{\boldmath$\epsilon$}}^{{\rm OTOC},f}~{\bf K}^f+ {\bf W}^{{\rm OTOC}, f} ~{\bf K}^f,\label{mas_froz-otoc}
\end{equation}
which  has similar form as the one derived in Ref.~\onlinecite{syzranov2018} for the case of a single reservoir in equilibrium. 
As already noticed in Ref.~\onlinecite{syzranov2018}, the master equation for the OTOC is basically the one for two copies of the density matrix.
In the case of a single reservoir and for a system without driving, Eq.~(\ref{master-otoc-fin}) is similar to implementing a forward and a backward evolution with the master equation for the density matrix as in Ref.~\onlinecite{gonza}. 
The adiabatic component can be calculated from
\begin{equation}
     \frac{\partial{\bf K}^f}{\partial{\bf X}} \dot{\bf X}= \frac{i}{\hbar} {\mbox{\boldmath$\epsilon$}}^{{\rm OTOC},f}~{\bf K}^a+ {\bf W}^{{\rm OTOC}, f} ~{\bf K}^a+ \mbox{\boldmath$\delta$}{\bf W}^{\rm OTOC}~ {\bf K}^f,\label{mas_ad-otoc}
\end{equation}
where the last term can be neglected using similar arguments to those presented as in the case of the adiabatic evolution for the density matrix. Also in the present case, we must take into account the contributions due to the changes of the matrix of Eq. (\ref{mas_froz-otoc}) and those corresponding to the changes in the eigenstates.  The corresponding solutions satisfy
\begin{multline}
\label{norm}
\sum_{mn}K_{mm,nn}^f=\left\langle{\cal O}_B(t_r){\cal O}_D(t_r)\right\rangle = K_{\infty}~~(\text{constant}),
\end{multline}
for the frozen components, and
\begin{equation}
\sum_{mn}K_{mm,nn}^a=0 ,
\end{equation}
for the adiabatic ones. 

\subsection{Currents}
Similarly, substituting Eqs. (\ref{eq:lam}) and (\ref{eq:overlam}) in the definition of the energy currents, we get 
\begin{multline}
\label{enercur_gen}
J_{\alpha}^{(E)}(t)=\sum_{m,n,u}\Big[\epsilon_{um}\; \tilde{W}^{um, f}_{mn,\alpha}(t)
\rho_{un}(t)  \\
-\epsilon_{nu}\; 
W^{nu, f}_{mn,\alpha}(t)
\rho_{mu}(t)\Big].
\end{multline}
Similarly, for the charge currents we obtain
\begin{equation}
\label{charcur_gen}
J_{\alpha}^{(c)}(t)= e\sum_{m,n,u}\Big[
 \tilde{V}^{um, f}_{mn,\alpha}(t) \rho_{un}(t) -
V^{nu, f}_{mn,\alpha}(t) \rho_{mu}(t)\Big],
\end{equation}
where
\begin{multline}\label{ratefch1}
V^{ju, (f)}_{ml,\alpha}= \lambda_{\alpha,ml}(t) \overline{\lambda}_{\alpha,ju}(t) \gamma^f_\alpha(\epsilon_{ju})/2 \\
-\overline{\lambda}_{\alpha,ml}(t)
\lambda_{\alpha,ju}(t) \tilde{\gamma}^f_\alpha(\epsilon_{uj})/2 ,
\end{multline}
\begin{multline}\label{ratefch2}
\tilde{V}^{ul, (f)}_{jm,\alpha}=\lambda_{\alpha,jm}(t) \overline{\lambda}_{\alpha,ul}(t) \tilde{\gamma}^f_{\alpha}(\epsilon_{ul})/2\\
-\overline{\lambda}_{\alpha,jm}(t) \lambda_{\alpha,ul}(t) \gamma^f_{\alpha}(\epsilon_{lu})/2.
\end{multline}

Notice that the charge current has been defined only for the fermionic case.

Moreover, currents are made up of a frozen and an adiabatic contributions [$J_{\alpha}^{(E),f}(t)$ and $J_{\alpha}^{(E),a}(t)$, respectively, for energy currents] coming from the respective terms of the density matrix.

\section{Examples}\label{exam}
The outcome of the previous sections is that the master equations describing the adiabatic dynamics of an open quantum system for the density matrix and the OTOC, as well as the currents, are completely defined by  the frozen rates in Eq.~(\ref{ratef1}) and (\ref{ratef2}). In order to calculate them, all we need is the unitary transformation diagonalyzing the instantaneous Hamiltonian of the system, the spectral function of the baths and the corresponding Bose-Einstein or Fermi-Dirac distribution functions. We illustrate the procedure for two simple examples.

\subsection{Qutrit}
\label{qutrit}
We  analyze the dynamics of a driven  {\em qutrit} --- a three-level system such as an atom  with a ground state and two excited states ---  attached to two bosonic reservoirs. The latter could, for instance, represent two electromagnetic environments to which the atom is coupled. We consider the following Hamiltonian for the driven three-level system
\begin{equation}
{\cal H}_{\rm S}(t)=\sum_{q=0}^2 E_{q}(t)\hat{\pi}_{qq}+w(t)\Big(\hat{\pi}_{12}+\hat{\pi}_{21}\Big),
\end{equation}
where $E_q(t)$, with $q=0,1,2$, are the energy levels relative to the ground state (0) and the two excited states (1 and 2). 
The inter-level coupling parameter $w(t)$ denotes the amplitude for, possibly, time-dependent transitions between the two excited states.
The consider the bath Hamiltonian described by Eq.~(\ref{Hb}) with $N_r=2$, $\hat{b}_{k\alpha}$ being bosonic operators for reservoir $\alpha=L,R$. As shown in Fig.~\ref{fig:sketch}, we fix the temperature of the two baths as $T_{\rm L}=T+\Delta T$ and $T_{\rm R}=T-\Delta T$. Moreover, we will consider Ohmic baths with linear dissipation relation spectral density
\begin{equation}
\Gamma_{\alpha}(\varepsilon)=\Upsilon_{\alpha}\,\varepsilon \,e^{{-\varepsilon}/{\epsilon_{c}}},~~~~\text{with~}\varepsilon>0,
\end{equation}
where $\epsilon_{c}$, is a  high frequency cut-off.
We assume that the left bath is connected to the qutrit through energy level $1$ and right bath is connected through energy level $2$, so that the contact Hamiltonian is given by
\begin{align}
{\cal H}_{\rm C}=&\sum_{k}V_{kL}\left(\hat{\pi}_{10}+\hat{\pi}_{01}\right)\left(b_{kL}^{\dagger}+b_{kL}\right)\nonumber \\
&+\sum_{k}V_{kR}\left(\hat{\pi}_{02}+\hat{\pi}_{20}\right)\left(b_{kR}^{\dagger}+b_{kR}\right).
\label{conqtrt}
\end{align} 
As detailed in Sec.~\ref{gf}, we first diagonalize  the system Hamiltonian with a suitable unitary transformation $\hat{U}(t)$ so that
\begin{equation}
\tilde{\mathcal{H}}_{\rm S}(t)
=\sum_{l=\pm}\varepsilon_{l}(t)\hat{\rho}_{ll}+\varepsilon_0(t)\hat{\rho}_{00},
\end{equation}
with the instantaneous eigenstates being $|0\rangle,~|-\rangle,~|+\rangle$, being the instantaneous eigenenergies $\varepsilon_{0}(t)=E_{0}(t)$ and
\begin{equation}
\varepsilon_{\pm}(t)=\left(\frac{E_{1}(t)+E_{2}(t)}{2}\right)\pm\frac{1}{2}\sqrt{\left(E_{1}(t)-E_2(t)\right)^{2}+4w(t)^{2}}.
\label{diagen}
\end{equation}
Moreover, the contact Hamiltonian in the instantaneous basis becomes
\begin{equation}
\tilde{{\cal H}}_{\rm C}=\sum_{k,\alpha}\sum_{l=\pm}V_{k\alpha}\Big(\lambda_{\alpha,0l}(t) \,\hat{\rho}_{0l}+\lambda_{\alpha,l0}(t)\,\hat{\rho}_{l0}\Big)\Big(b_{k\alpha}+b_{k\alpha}^{\dagger}\Big),
\label{qutritcont}
\end{equation}
where 
\begin{eqnarray}\label{lambqut}
\lambda_{L,0+}(t)&=&-\lambda_{R,0-}(t)=\cos\theta(t)/2, \nonumber \\
\lambda_{L,0-}(t)&=&\lambda_{R,0+}(t)=\sin\theta(t)/2, 
\end{eqnarray}
with $\theta(t) = \tan^{-1} \left(\frac{2w(t)}{E_1(t) -E_2(t)}\right)$ and $\lambda_{\alpha,0l}(t)=\lambda_{\alpha,l0}(t)$. In the present problem
$\overline{\lambda}_{\alpha,ml}(t)=\lambda_{\alpha,lm}(t)$

\subsubsection{Density matrix}
Given Eqs.~(\ref{diagen}) we immediately have the component of the frozen kernel  ${\mbox{\boldmath$\epsilon$}}$ in Eq.~(\ref{mas_froz}). On the other hand, given Eqs.~(\ref{lambqut}), 
 we readily get the rates defined in Eqs.~(\ref{ratef1}) and (\ref{ratef2}), which completes all the information about the elements of the frozen master equation in Eq.~(\ref{mas_froz}).

\begin{figure}[!thb]
	\centering
  \includegraphics[width=0.97\columnwidth]{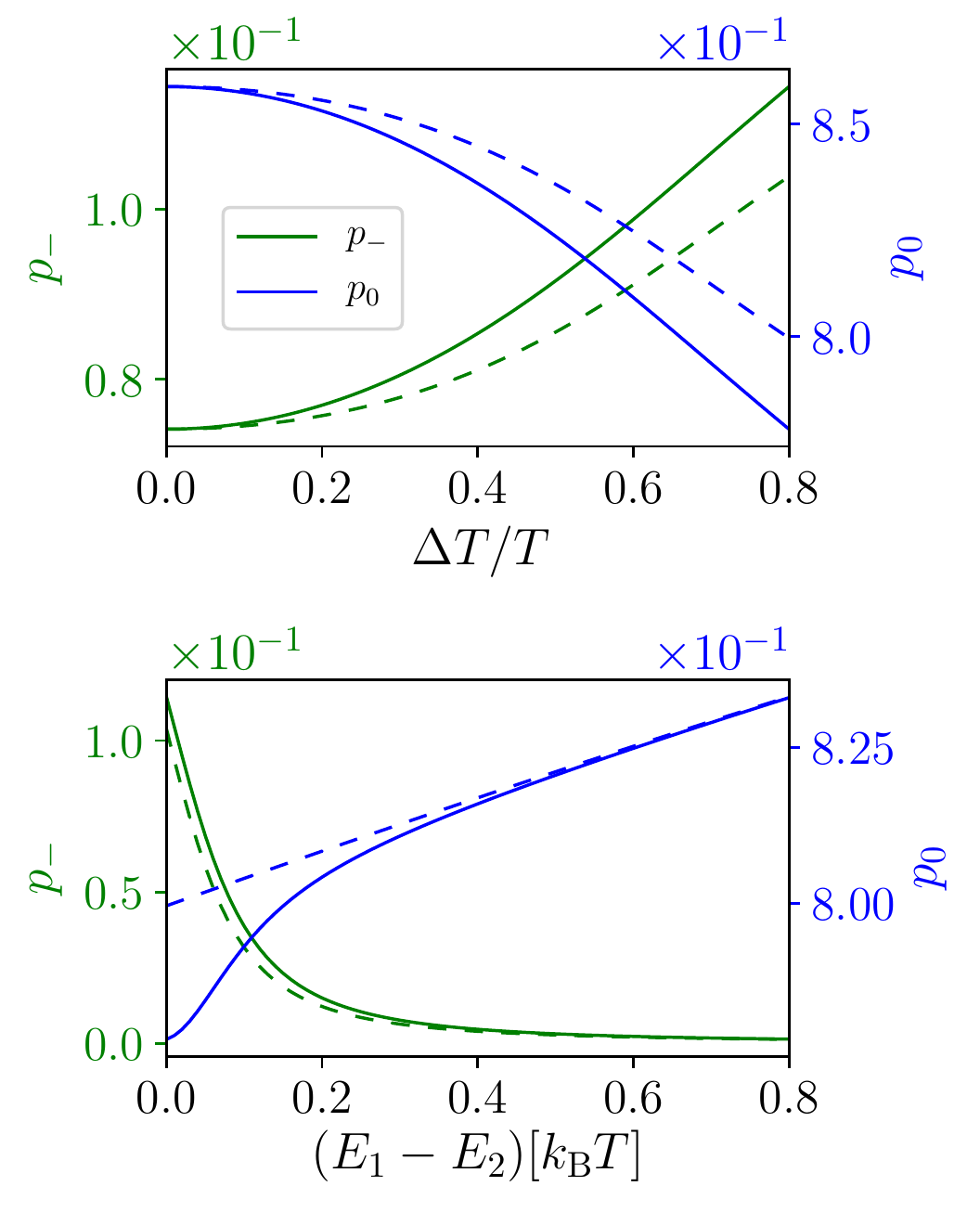}
\caption{Qutrit: population probabilities in the absence of driving as functions of $\Delta T$ for fixed energy splitting $E_1-E_2=0$ (top panel), and as functions of level splitting for $\Delta T= 0.8 T$ (bottom panel).
Green lines refers to $p_-$ (with axis on the left) and blue lines refers to $p_0$ (with axis on the right).
Notice the two scales on the left and right axis are different.
Solid (dashed) lines result from the solution of the RE (QME).
Parameters values are: $\Upsilon_L=\Upsilon_R=0.2$, $\epsilon_{C}=100~k_{B}T$,  $w=0.05~k_{B}T$, $E_1+E_2=5~k_{B}T $, and $E_0=0$.}
	\label{fig:popvsereldT}
\end{figure}

We first consider the particular case of the frozen Hamiltonian, corresponding to fixed values of $E_l$ and $w$,
 and we analyze the effect of coherence by comparing the outcomes of the full quantum master equation (QME)  with those obtained from the rate equation (RE). The latter corresponds to solving the equation for the diagonal elements only. 
In Fig.~\ref{fig:popvsereldT} we plot the populations $p_0$ (blue lines) and $p_-$ (green lines) of the states $|0\rangle$ and $|-\rangle$, respectively, as functions of $\Delta T$ (top panel) and energy level splitting $E_1-E_2$ (bottom panel).
Solid lines result from the solution of the RE, while dashed lines from the solution of the QME.
The top panel in Fig.~\ref{fig:popvsereldT} shows that the effect of coherence on the populations is absent for $\Delta T=0$ (where the overall system is at equilibrium) and leads to an important contribution  for large values of $\Delta T$.
The bottom panel of Fig.~\ref{fig:popvsereldT}, highlights the relevant energy scale for which the coherence plays a significant role. Concretely, we see that this is the case when the level splitting is small compared to $k_B T$. In the figure this corresponds to values $E_1-E_2 < 0.5$ $k_BT$, in which range,
for this choice of parameters, the gap between bonding and anti-bonding energy levels $\Delta=\varepsilon_+-\varepsilon_-$ is smaller than the energy scale $k_B\Delta T$~\cite{cuetara}.
\begin{figure}[!thb]
	\centering
  \includegraphics[width=0.97\columnwidth]{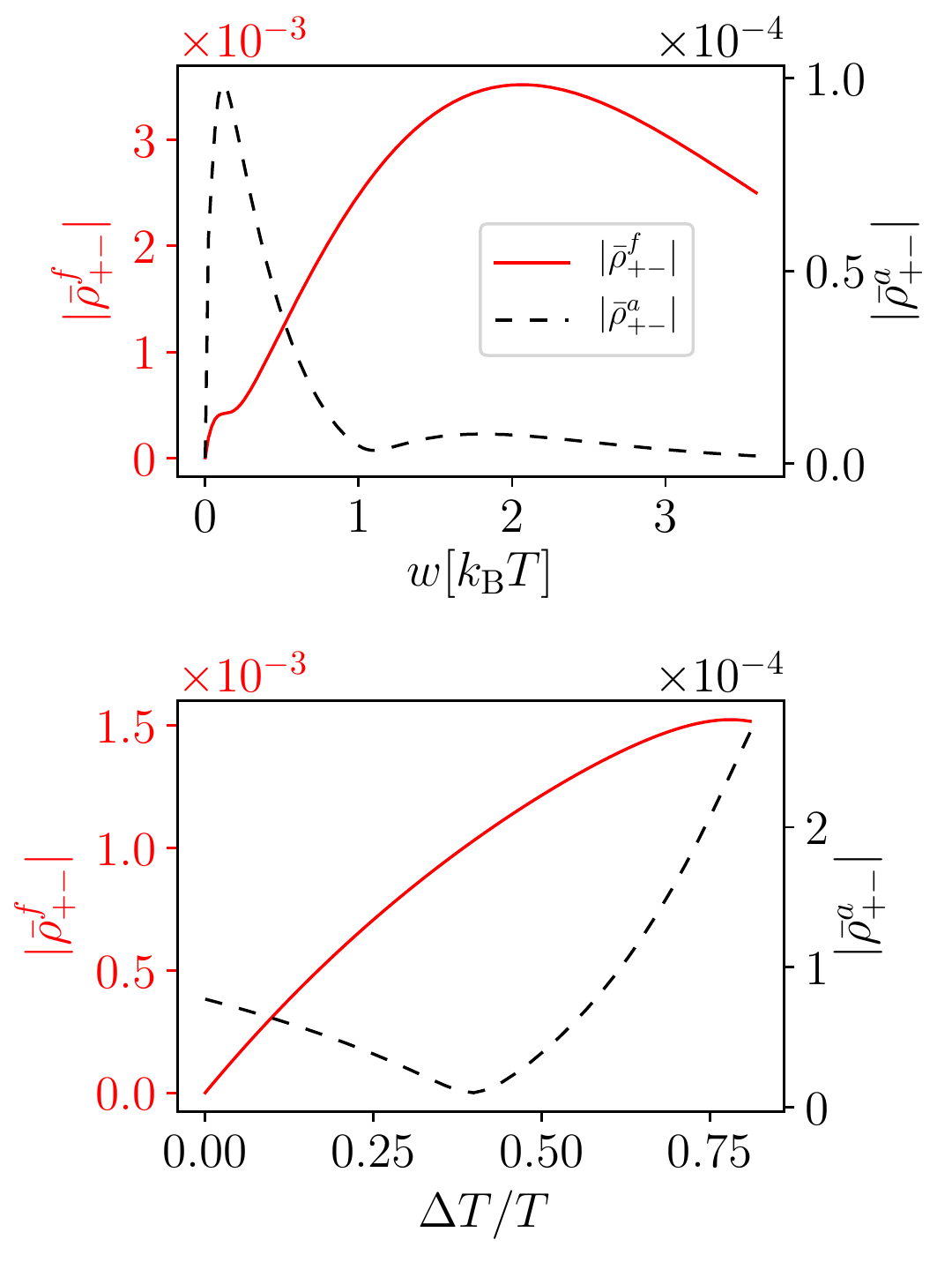}
\caption{Qutrit: frozen ($f$) and adiabatic ($a$) contributions to the period-averaged coherences $\bar{\rho}_{+-}$ as functions of the inter-level coupling $w$ for fixed $\Delta T = 0.5 T$ (top panel), and as functions of $\Delta T$ for fixed $w=0.5~k_{B}T$ (bottom panel).
Red solid lines refer to the absolute value of the frozen contribution (with axis on the left) and black dashed lines refer to the absolute value of the adiabatic contribution (with axis on the right).
Parameters values are: $\Upsilon_L=\Upsilon_R=0.2$, $\epsilon_{c}=30~k_{B}T$, $E_1(t)=2 k_{\rm B}T+4k_{\rm B}T \cos{(\Omega t+\frac{\pi}{2})}$, $E_0=0$ and $E_2(t)=0.5k_{\rm B}T \cos{(\Omega t)}$. }
\label{fig:offvsdT}
\end{figure}

Let us now assume that the system is driven by modulating the parameters according to the following scheme: $E_1(t)+E_2(t)=E_{\rm av}+\delta \varepsilon \cos(\Omega t+\phi)$ and $E_1(t)-E_2(t)=E_{\rm rel}+\delta \bar{\varepsilon} \cos(\Omega t)$, while $w$ is time-independent. 
In the present case, we solve the adiabatic master equation of Eq.~(\ref{mas_ad}) along with the frozen one in Eq.~(\ref{mas_froz}), which also accounts for the gauge potential term.
In Fig.~\ref{fig:offvsdT} we plot the absolute value of the period-averaged frozen and adiabatic contributions to the coherences, $\bar{\rho}^{f}_{+-}$ and $\bar{\rho}^{a}_{+-}$, respectively,
as functions of $w$ (top panel) and $\Delta T$ (bottom panel).
Red solid lines refer to the frozen contribution (with values on the left axis) and black dashed lines refer to the adiabatic contribution (with value on the right axis).
The top panel of Fig.~\ref{fig:offvsdT} shows that frozen and adiabatic components of the coherence display different behaviors as a function of the inter-level coupling, while their absolute values differ by more than one order of magnitude.
In particular, they are zero at $w=0$, since no coupling is present, and are suppressed at large $w$, since the gap $\Delta$ gets larger than the energy scale $k_B\Delta T$. However, they present a maximum at different values of $w$, namely at about $w=2 k_BT$, for the frozen component, and at about $w=0.2 k_BT$, for the adiabatic component.
The bottom panel of Fig.~\ref{fig:offvsdT} shows that the two components of the coherence $\bar{\rho}_{+-}$ behave differently as functions of $\Delta T$. While the frozen component vanishes for $\Delta T=0$ and thereafter increases, the adiabatic component first decreases (starting from a finite value at $\Delta T=0$), reaching a minimum at $\Delta T\simeq 0.4T$ and thereafter increasing.

\subsubsection{OTOC}
\begin{figure}[!htb]
	\centering
  \includegraphics[width=0.97\columnwidth]{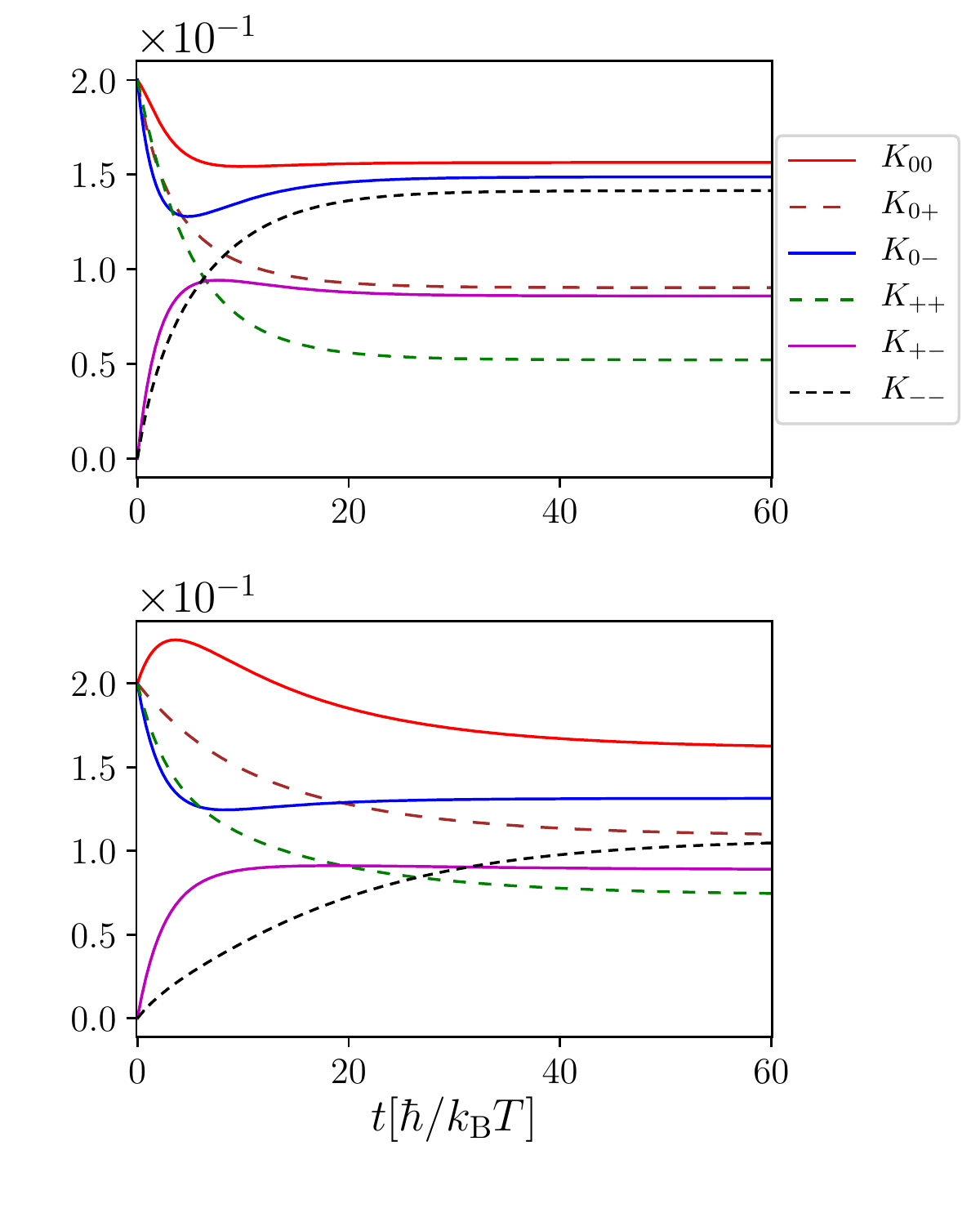}
\caption{Qutrit: time variations of OTOCs for different values of thermal bias $\Delta T=0$ (upper panel) and $\Delta T=0.98T$ (lower panel) taking only the diagonal terms of the projection operators and $\Upsilon_L=0.1$, $\Upsilon_R = 0.2$,
$w = 0.2 k_{\rm B}T$, $E_0=0$, $E_{1}-E_{2} = 0.3k_{\rm B}T$, $E_{1}+E_2 = 0.6k_{\rm B}T$,  $ \epsilon_{\rm c} = 20 k_{\rm B}T$, with initial conditions $K_{00}(0)=0.2$, $K_{0+}(0)=0.2$, $K_{0-}(0)=0.2$, $K_{+0}(0)=0.2$, $K_{++}(0)=0.2$ and the normalization condition $\sum_{ll'}K_{ll'}(t)=1$.}
	\label{fig:otoc_diag}
\end{figure}

The different components of the quantity $K_{ll'}(t)=K_{ll,l'l'}(t)$, corresponding to the solution of Eq.~(\ref{master-otoc-fin}) without driving, are shown in Fig.~\ref{fig:otoc_diag} for 
different values of the initial conditions.
Here the gauge term of the master equation (\ref{mas_ad-otoc}) does not play any role since  $K_{ll'}$  correspond to the ``rate" component of the OTOC.
Because of the normalization conditions in Eq.~(\ref{norm}), we have normalized the operators ${\cal O}(t)$ such that $\sum_{ll'}K_{ll'}(t)=1$.
We recall that OTOCs have been suggested as useful quantities to characterize scrambling dynamics in many-body systems. In non-integrable Hamiltonians, OTOC's are expected to grow as a function of time \cite{kitaev,maldacena,ioffe,sachdev}. In systems coupled to thermal baths these correlations stabilize after some time \cite{syzranov2018,gonza} and tend to the asymptotic limit determined by Eq.~(\ref{norm}). This can be appreciated in
the evolution shown in Fig.~\ref{fig:otoc_diag}. In the bottom panel of the same figure we show the evolution of the OTOCs for the same parameters and the same initial conditions shown in the top panel, under the presence of a thermal bias between the two reservoirs. Overall, we see a similar behavior as in the case of equilibrium reservoirs shown in the top panel. However, we can notice that the time to reach the asymptotic limit is larger and we can see that in some components there is an enhancement with respect to the equilibrium case. 

\begin{figure}[!htb]
	\centering
  \includegraphics[width=0.97\columnwidth]{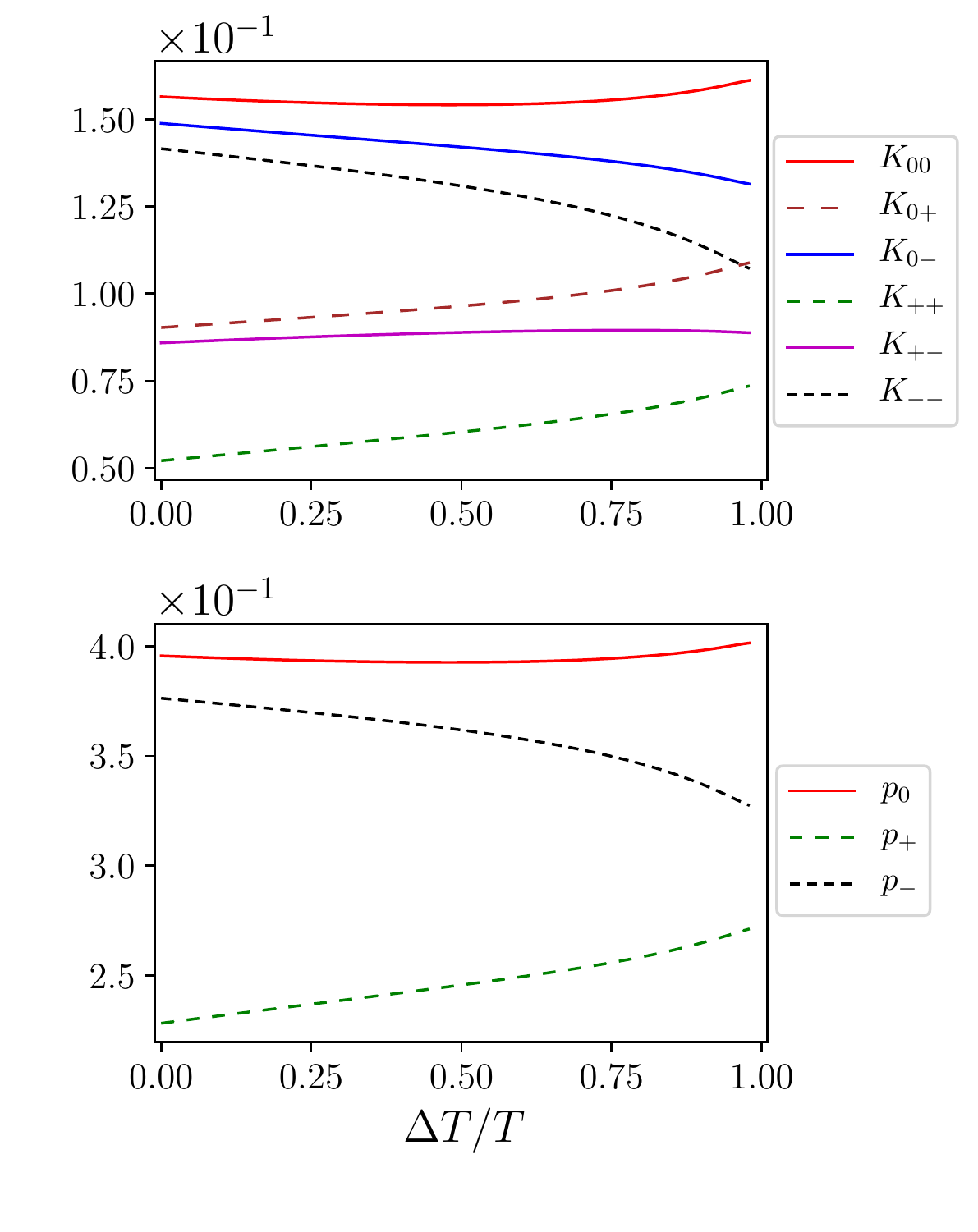}
  \caption{Qutrit: steady state values of OTOCs as a function of thermal bias for the time independent case taking only the diagonal terms of the projection operators. The parameters are the same as in Fig.~\ref{fig:otoc_diag}. As a reference, we also show in the bottom panel the populations of the different levels for the same parameters.}
  \label{fig:otoc-statio}
  \end{figure}

In Fig.~\ref{fig:otoc-statio} we can see the behavior of the stationary values solution of Eq.~(\ref{mas_froz-otoc}) for the same parameters as in Fig.~\ref{fig:otoc_diag} in the non-equilibrium regime. For some components we can see an enhancement of the OTOCs as the temperature bias $\Delta T$ increases.
The behavior of
$K_{00},~K_{--},~K_{++}$ is correlated with the 
behavior of the populations of the different levels, as shown in the bottom panel of Fig.~\ref{fig:otoc-statio}. The reason is that the thermal bias generates relative changes in the populations, which are accounted for by the non-equilibrium features of the OTOCs.

\begin{figure}[!htb]
	\centering
    \includegraphics[width=0.97\columnwidth]{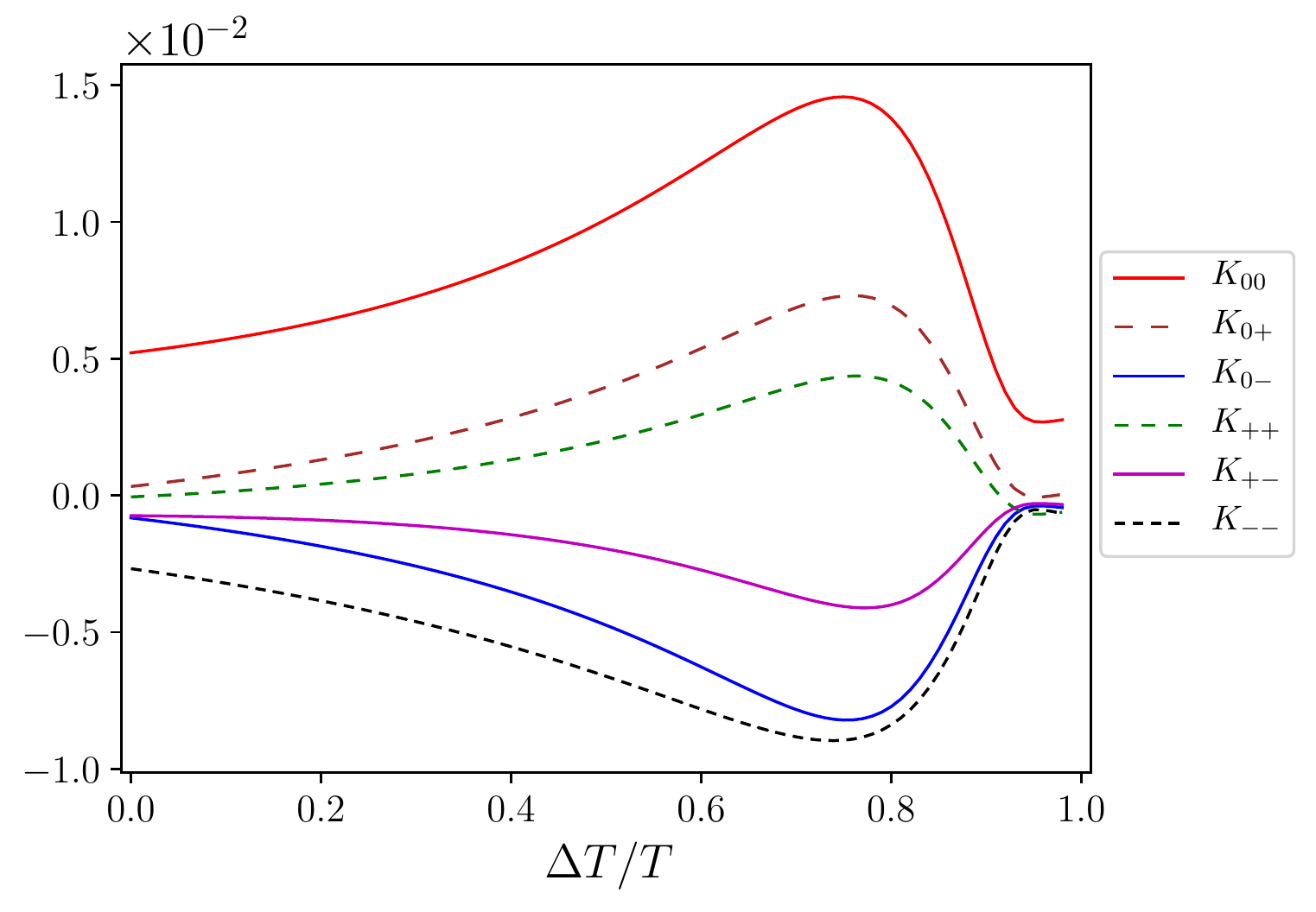}
\caption{Qutrit: Adiabatic contribution to the OTOCs as a function of thermal bias under adiabatic driving taking only the diagonal terms of the projection operators and $\Upsilon_L=0.2$, $\Upsilon_R = 0.2$,
$w = 0.05 k_{\rm B}T$, $\epsilon_{C}=20~k_{B}T$, $E_0=0$, $E_1(t)+E_2(t)=2k_{\rm B}T+2k_{\rm B}T\cos{(\Omega t+\frac{\pi}{2})}$,  $E_1(t)-E_2(t)=0.5k_{\rm B}T+0.5k_{\rm B}T\cos{(\Omega t)}$, $\hbar\Omega=0.01 k_{\rm B}T$, and the normalization condition $\sum_{ll'}K_{ll'}^f=1$.}
	\label{fig:otoc_diag3}
\end{figure}

  Finally, the effect of driving treated in the framework of the adiabatic approximation is illustrated in Fig.~\ref{fig:otoc_diag3}. We recall that these are corrections to the frozen components shown in Fig.~\ref{fig:otoc-statio}.
 Interestingly, they are positive for the components $K_{00},~K_{++},~K_{0+}$, which are the ones that grow with $\Delta T$ in the frozen case, while they are negative for the ones that decrease.

\subsection{Coupled quantum dots}
\label{Sec_cqd}
In this section we analyze now a fermionic driven system consisting of a pair of coupled quantum dots (QDs).\cite{janine1}
For simplicity, we focus on the case with infinite intra-dot Coulomb repulsion, which limits the occupation to, at the most, one electron per quantum dot, and we assume spinless fermions.
 
Concretely, we consider the following Hamiltonian for a pair of coupled single-level quantum dots 
\begin{align}
{\cal H}_{\rm S}(t)= E_{1}(t)~\hat{a}_{1}^{\dagger}\hat{a}_{1}+E_{2}(t)~\hat{a}_{2}^{\dagger}\hat{a}_{2}+\nonumber\\
+w(t)~(\hat{a}_{1}^{\dagger}\hat{a}_{2}+\hat{a}_{2}^{\dagger}\hat{a}_{1})+U\hat{n}_{1}\hat{n}_{2}, 
\label{sysqd}
\end{align}
where $\hat{a}_{j}$ and $\hat{a}_{j}^{\dagger}$ are, respectively, the annihilation and creation operators for fermions in the quantum dot $j=1,2$ and $\hat{n}_j=\hat{a}_{j}^{\dagger}\hat{a}_{j}$.
The time-dependent parameters are the QDs' energy levels $E_{1}(t)$ and $E_{2}(t)$, and the hopping element $w(t)$ between the two QDs, while $U$ is the inter-dot Coulomb interaction.
The bath Hamiltonian is given by Eq.~(\ref{Hb}) with $N_r=2$, $\hat{b}_{k\alpha}$ being fermionic operators for reservoir $\alpha=L,R$.
Moreover, we assume a characterless spectral density, namely $\Gamma_\alpha(\epsilon)=\Gamma_\alpha$, independent of energy.
The contact Hamiltonian is given by
\begin{equation}
{\cal H}_{\rm C}=\sum_{k}V_{kL}\hat{b}_{kL}^{\dagger}\hat{a}_{1}+\sum_{k}V_{kR}\hat{b}_{kR}^{\dagger}\hat{a}_{2}+ h.c.,
\label{conqd}
\end{equation}
so that each QD is connected only to one reservoir.

The Hilbert space of the double-dot system is composed of the following four occupation states: $|0\rangle$ (empty), $|1\rangle= \hat{a}^{\dagger}_1|0\rangle$ (single occupancy, left QD),  $|2\rangle= \hat{a}^{\dagger}_2|0\rangle$ (single occupancy, right QD) and $|d\rangle = \hat{a}^{\dagger}_1\hat{a}^{\dagger}_2|0\rangle$ (double occupancy).
%
%
The diagonalyzed system Hamiltonian reads
\begin{equation}
\tilde{\mathcal{H}}_{\rm S}(t)=
\sum_{l=0,\pm,d}\varepsilon_{l}(t)\hat{\rho}_{ll},
\end{equation}
where $\varepsilon_{\pm}(t)$ are given by Eq.~(\ref{diagen}), $\varepsilon_0=0$ and $\varepsilon_d(t) = U+ \varepsilon_1(t)+\varepsilon_2(t)$.
The contact Hamiltonian becomes
\begin{equation}
\tilde{\mathcal{H}}_{\rm C} = \sum_{k,\alpha}\sum_{l=\pm}V_{k\alpha}\Big[\lambda_{\alpha,0l}(t)\,\hat{b}_{k\alpha}^\dagger \hat{\rho}_{0l}+\lambda_{\alpha,dl}(t)\,\hat{b}_{k\alpha}^\dagger \hat{\rho}_{ld}+ h.c.\Big],
\end{equation}
where $\lambda_{L,0+}(t)=-\lambda_{R,0-}(t)=\lambda_{R,+d}(t)=\lambda_{L,-d}(t)$ are the same as in Eq. (\ref{lambqut}), while $\lambda_{\alpha,ll^{\prime}}(t)= \lambda_{\alpha,l^{\prime}l}(t)$.
As before, the full adiabatic master equation for diagonal and off-diagonal terms of the density matrix and the OTOC can be obtained after calculating Eqs.~(\ref{ratef1}) and (\ref{ratef2}) following Sec.~\ref{fame}.
\begin{figure}[!htb]
	\centering
  \includegraphics[width=0.97\columnwidth]{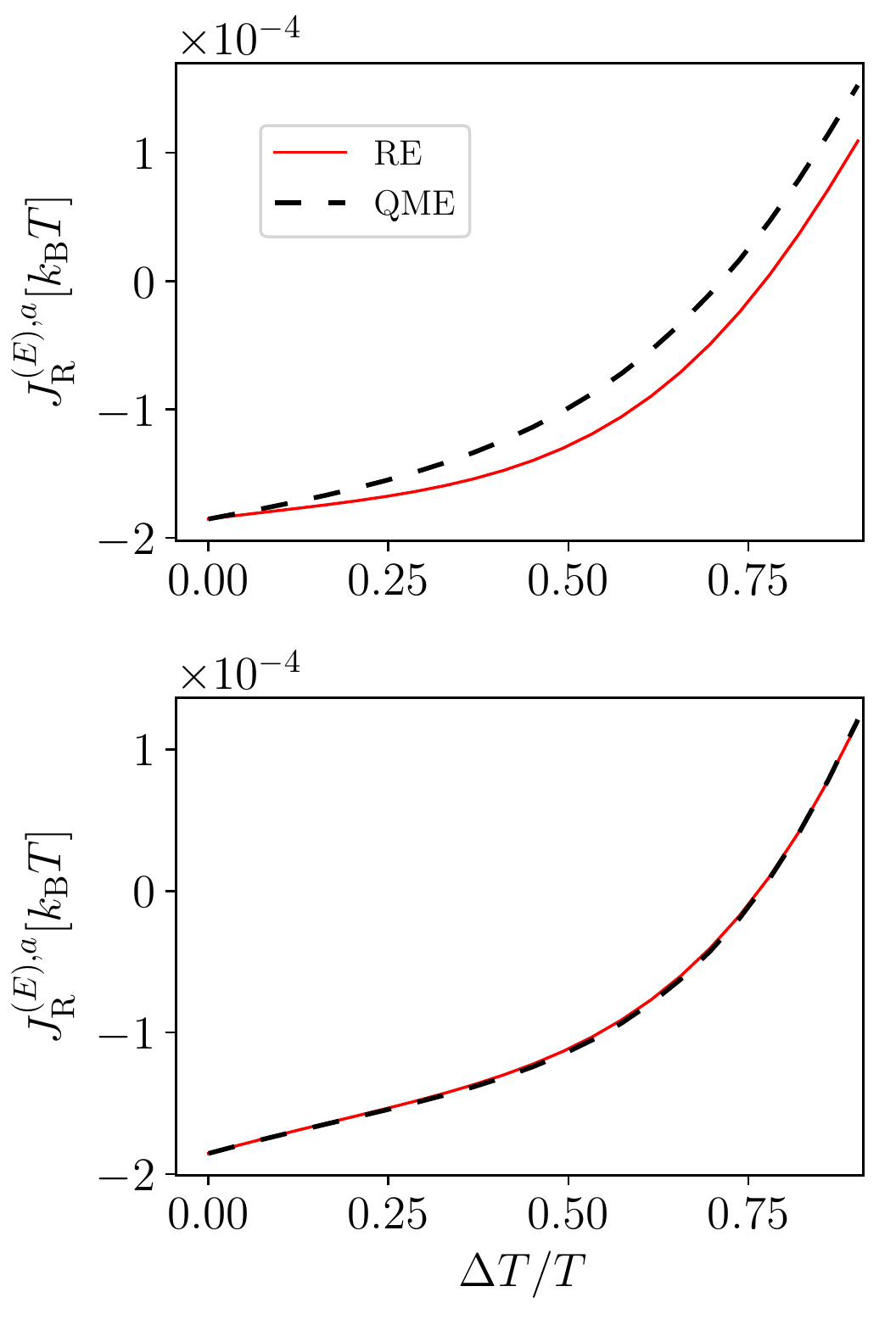}
\caption{Adiabatically-pumped energy current in the right reservoir averaged over one period relative to the qutrit system (top panel) and to the coupled QD system (bottom panel) as a function of a $\Delta T$. Solid red (dashed black) lines results from the solution of the RE (QME).
Parameters values are:
$w = 0.2 k_{\rm B}T$, $\epsilon_{\rm c}=100k_{\rm B}T$, $E_1(t)=2k_{\rm B}T+2k_{\rm B}T\cos{(\Omega t-\frac{\pi}{2})}$, $-E_2(t)=k_{\rm B}T+k_{\rm B}T\cos{(\Omega t)}$, $\hbar\Omega=0.01 k_{\rm B}T$; for qutrit $\Upsilon_L= \Upsilon_R = 0.2$ and for coupled quantum dots $\Gamma_L= \Gamma_R = 0.2 k_{\rm B}T$, $U=0$.}
	\label{fig:curvsdTco}
\end{figure}
Since the kernel ${\bf W}$ is very similar to the qutrit case, the behavior of the matrix elements $\rho_{jl}$ is qualitatively similar to what is shown in Figs.~\ref{fig:popvsereldT} and \ref{fig:offvsdT}.
 
Our aim now is to calculate the energy currents flowing through the system between the reservoirs as a consequence of the combined effect of the thermal bias and the ac-driving. We consider
a modulation in time of the parameters $E_1(t)$ and $E_2(t)$ according to the scheme presented in Sec.~\ref{qutrit}, while taking $w$ time-independent.
We notice that, in the adiabatic regime, asymmetric coupling (with respect to the reservoirs) is a necessary condition to obtain a net pumping of energy over a period when $\Delta T=0$~\cite{adiageo}.
In Fig.~\ref{fig:curvsdTco} we plot the adiabatically-pumped energy current (averaged over a period) flowing into the right reservoir ${\bar{J}}^{(E),a}_R=\Omega/(2\pi)\int_0^{2\pi/\Omega}dt~J^{(E),a}_R(t)$ as a function of the $\Delta T$ for both systems, qutrit and coupled QD, corresponding, respectively, to the upper and lower panels. It is interesting to analyze here the role of the coherences in evaluating the currents. 
The solid red lines result from the solution of the RE, in the absence of coherence effects, while the dashed black lines result from the solution of the QME.
The fact that the value of ${\bar{J}}^{(E),a}_R$ is negative (in a range of values of $\Delta T$) means that the energy current is exiting the right, cold reservoir,
so that the system works as a refrigerator.
Interestingly, in both cases we find that the presence of coherence decreases the absolute value of the energy current, thus suppressing the refrigeration effect\cite{brandner,brandner2}. 
The effect is more pronounced for the qutrit than for the coupled QD case.

\section{Summary and conclusions} \label{conclu}
We have presented a derivation of the quantum master equation ruling the adiabatic dynamics of a driven system weakly coupled to non-equilibrium reservoirs. The formalism applies to any Hamiltonian system
with finite dimension of its Hilbert space at which a slowly varying time-dependent perturbation is applied and weakly coupled to fermionic or bosonic baths. Our derivation includes the equations for the dynamics of the reduced density matrix of the finite-size system, the currents between the system and the reservoirs and the out-of-time-order correlation (OTOC) functions. 

We have illustrated the application of the formalism with two examples: a qutrit coupled to two bosonic baths and two coupled quantum dots attached to fermionic baths. In both cases, a time-periodic perturbation with low frequencies, consistent with the adiabatic regime, and a temperature bias were considered. We showed the relevance of the off-diagonal terms of the density matrix (coherences) in the far-from equilibrium situations, corresponding to large temperature differences between reservoirs. We have also analyzed the steady state and adiabatic solutions of the OTOC. 

The present formalism may represent a useful tool to analyze the dynamics of the OTOC in other Hamiltonian systems coupled to reservoirs in non-equilibrium scenarios.

\section{Acknowledgements}
We thank Paolo Abiuso, Mart\'{\i} Perarnau-Llobet, Janine Splettstoesser and Pablo Terr\'en Allonso for useful duscusions. 
LA acknowledges support from PIP-2015-CONICET,  PICT-2017, PICT-2018, Argentina, Simons-ICTP-Trieste associateship, and the Alexander von Humboldt Foundation, Germany. We thank the support of the CNR-CONICET cooperation program ``Energy conversion in quantum, nanoscale, hybrid devices'', as well as the hospitality of the Dahlem  Center for Complex Quantum Systems, Berlin and the International Center for Theoretical Physics, Trieste.

\appendix
\section{Evaluation of the mean values}
\label{mean}
\begin{figure}[!thb]
\centering
\includegraphics[width
=\columnwidth]{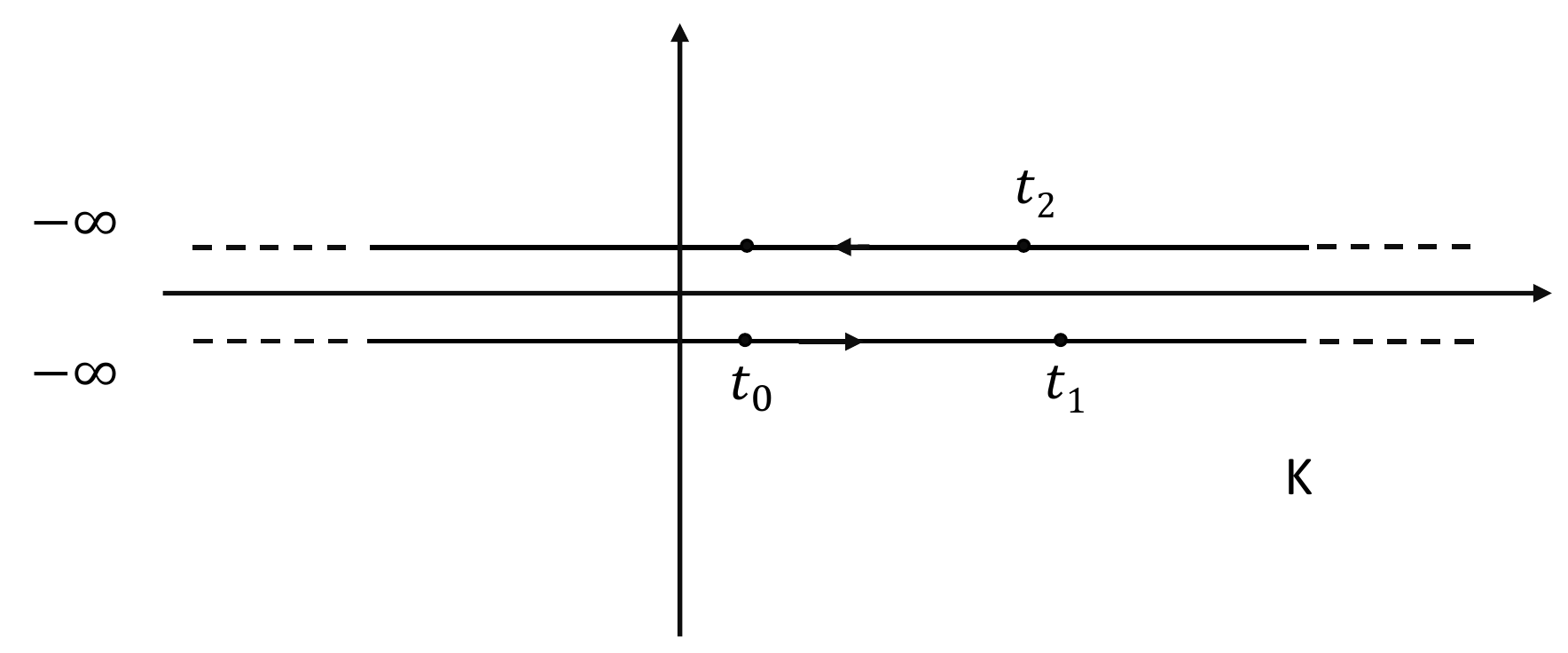}
\caption{The Keldysh contour.}
\label{fig:contour}
\end{figure}
In this section we will evaluate the mean values entering in Eq.~(\ref{master1}). The mean values will be calculated perturbatively up to first order in the coupling strength $V_{k\alpha}$, starting with the time-ordered correlator
\begin{align}
&i\left\langle T_{K} \hat{b}_{k\alpha}^\dagger (t')\hat{\rho}_{mj}(t)\right\rangle\nonumber \\
&\approx\int_{K}dt_1 \left\langle T_{K} \left[\tilde{\mathcal{H}}_{C}^{H}(t_1)\hat{b}_{k\alpha}^\dagger (t')\hat{\rho}_{mj}(t)\right]\right\rangle\nonumber \\
&= V_{k\alpha}\sum_{uv}\overline{\lambda}_{\alpha,uv}\int_K dt_1 \Big\langle T_{K}\Big[\hat{\rho}_{uv}(t_1)\hat{b}_{k\alpha}(t_1)\nonumber \\
&\hspace{0.5\columnwidth}\hat{b}_{k\alpha}^\dagger(t')\hat{\rho}_{mj}(t)\Big]\Big\rangle .
\end{align}
We can deform the contour $K$ into a pair of contours such that $K_1$ goes from $-\infty$ to $t$ and back to $-\infty$ and $K_2$ from $-\infty$ to $t^{\prime}$ and back to $-\infty$.
We can write
\begin{equation}
\int_{K_1}=\int_{-\infty}^t+\int_t^{-\infty};~\int_{K_2}=\int_{-\infty}^{t'}+\int_{t'}^{-\infty},
\end{equation}
\ \\
such that
\begin{widetext}
\begin{multline}
\label{mixxx}
i\left\langle\hat{b}_{k\alpha}^\dagger (t')\hat{\rho}_{mj}(t)\right\rangle  \approx\\
 V_{k\alpha}\sum_{uv}\overline{\lambda}_{\alpha,uv}\bigg[\left[\int_{-\infty}^t dt_1 \Big\langle \hat{b}_{k\alpha}^\dagger(t')\hat{\rho}_{mj}(t)\hat{\rho}_{uv}(t_1)\hat{b}_{k\alpha}(t_1)\Big\rangle+\int_{t}^{-\infty}dt_1 \Big\langle \hat{b}_{k\alpha}^\dagger(t')\hat{\rho}_{uv}(t_1)\hat{b}_{k\alpha}(t_1)\hat{\rho}_{mj}(t)\Big\rangle\right]  \\
+\left[\int_{-\infty}^{t^\prime} dt_1 \Big\langle\hat{b}_{k\alpha}^\dagger(t')\hat{\rho}_{uv}(t_1)\hat{b}_{k\alpha}(t_1)\hat{\rho}_{mj}(t)\Big\rangle+\int_{t^\prime}^{-\infty}dt_1 \Big\langle \hat{\rho}_{uv}(t_1)\hat{b}_{k\alpha}(t_1)\hat{b}_{k\alpha}^\dagger(t')\hat{\rho}_{mj}(t)\Big\rangle\right]\bigg].
\end{multline}
\end{widetext}
For the mixed lesser Green's function defined in Eq.~(\ref{gmix}), using Wick's theorem Eq.~(\ref{mixxx}) can be re-written as
\begin{align}
&G_{mj,k\alpha}^<(t,t') \nonumber \\
&\approx \int_{-\infty}^\infty dt_1 V_{k\alpha}\sum_{uv}\overline{\lambda}_{\alpha,uv}(t)\Big[g_{mj,vu}^r(t,t_1)g_{k\alpha}^<(t_1,t')\nonumber \\
&\hspace{0.5\columnwidth}+g_{mj,vu}^<(t,t_1)g_{k\alpha}^a(t_1,t')\Big],
\end{align}
where the definition for the system Green's functions are given in Eqs.~(\ref{gsys}) and (\ref{gsys2}).
The lesser and greater Green's function for the baths are defined as
\begin{align}
g_{k\alpha}^<(t_1,t_2)&=\pm i \left\langle b_{k\alpha}^\dagger (t_2)b_{k\alpha}(t_1)\right\rangle\nonumber \\
g_{k\alpha}^>(t_1,t_2)&=-i \left\langle b_{k\alpha} (t_1)b_{k\alpha}^\dagger(t_1)\right\rangle,
\end{align}
where the upper sign applies to fermionic reservoirs and the lower sign is for bosonic reservoirs. The corresponding retarded and advanced Green's functions can be obtained using a similar relation as in Eq.~(\ref{gsys2}).

\section{Frozen and adiabatic components of lesser Green's function}
\label{app:den_ada}
The lesser Green's function is given by:
\begin{equation}
g_{lj,vu}^<(t_1,t_2)=\pm i \left\langle \hat{\rho}_{uv}(t_2)\hat{\rho}_{lj}(t_1)\right\rangle.
\end{equation}
Writing in terms of evolution operators,
\begin{multline}
g_{lj,vu}^<(t_1,t_2)=\pm i \Big\langle \hat{T}_K e^{{i}/{\hbar}\int_{t_0}^{t_2}\tilde{\cal{H}}_S(t')dt'}\hat{\rho}_{uv}\\
e^{{-i}/{\hbar}\int_{t_1}^{t_2}\tilde{\cal{H}}_S(t')dt'}\hat{\rho}_{lj}e^{{-i}/{\hbar}\int_{t_0}^{t_1}\tilde{\cal{H}}_S(t')dt'}\Big\rangle
\end{multline}
Using $\tilde{\cal{H}}_S(t')=\tilde{{\cal H}}^f_{\rm S}+ \delta \tilde{{\cal H}}_{\rm S}(t^{\prime})$ along with Eq.~(\ref{eq:frogless}), we obtain
\begin{multline}
g_{lj,vu}^<(t_1,t_2)=\pm i \Big\langle \hat{T}_K e^{{i}/{\hbar}\int_{t_0}^{t_2}\delta\tilde{\cal{H}}_S(t')dt'}\hat{\rho}_{uv}^{f}(t_2)\\
e^{{-i}/{\hbar}\int_{t_1}^{t_2}\delta\tilde{\cal{H}}_S(t')dt'}\hat{\rho}_{lj}^{f}(t_1)e^{{-i}/{\hbar}\int_{t_0}^{t_1}\delta\tilde{\cal{H}}_S(t')dt'}\Big\rangle
\end{multline}

For the contour shown in Fig.~\ref{fig:contour}, where contour $K$ goes from $t_0\rightarrow t_1\rightarrow t_2 \rightarrow t_0$, we can write
\begin{equation}
\label{11}
g_{lj,vu}^<(t_1,t_2)=\pm i \left\langle \hat{T}_K e^{{-i}/{\hbar}\int_K\delta\tilde{\cal{H}}_S^{f}(t')dt'}\hat{\rho}_{uv}^{f}(t_2^-)\hat{\rho}_{lj}^{f}(t_1^+)\right\rangle,
\end{equation}
where we used $\delta\tilde{\cal{H}}_S\equiv \delta\tilde{\cal{H}}_S^f$ considering the driving to be slow enough. One other simplification entailed by slow driving is that one can Taylor expand the exponential in Eq.~(\ref{11}), obtaining
\begin{equation}
e^{{-i}/{\hbar}\int_K\delta\tilde{\cal{H}}_S^{f}(t')dt'}\approx 1{-i}/{\hbar}\int_K\delta\tilde{\cal{H}}_S^{f}(t')dt'
\end{equation}
such that
\begin{multline}
g_{lj,vu}^<(t_1,t_2)=\pm i \left\langle \hat{T}_K \hat{\rho}_{uv}^{f}(t_2^-)\hat{\rho}_{lj}^{f}(t_1^+)\right\rangle \\
\pm  {1}/{\hbar} \left\langle \hat{T}_K \int_K\delta\tilde{\cal{H}}_S^{f}(t')dt'\;\hat{\rho}_{uv}^{f}(t_2^-)\hat{\rho}_{lj}^{f}(t_1^+)\right\rangle,
\end{multline}
where the first term on the r.h.s. is the frozen contribution to the lesser Green's function,
\begin{eqnarray}
g_{lj,vu}^{<,f} (t_1,t_2) &= &\pm i  \delta_{lv} \left\langle\hat{\rho}_{uj}^f(t_1)\right\rangle e^{i \epsilon_{uv}(t_2-t_1)} \nonumber \\
&=& \pm i  \delta_{lv} \left\langle\hat{\rho}_{uj}^f(t_2)\right\rangle e^{i \epsilon_{jv}(t_2-t_1)},
\end{eqnarray}
whereas the second term gives higher order contributions,
\begin{multline}
\delta g_{lj,vu}^{<,f}(t_1,t_2)\\
=\pm  {1}/{\hbar}\sum_{k=1}^{N} \left\langle \hat{T}_K \int_K \xi_k(t')\hat{\rho}_{kk}^{f}(t')dt'\;\hat{\rho}_{uv}^{f}(t_2^-)\hat{\rho}_{lj}^{f}(t_1^+)\right\rangle,
\end{multline}
where we used the second equation of Eq.~(\ref{exp}). Expanding over the Keldysh contour we get
\begin{multline}
\delta g_{lj,vu}^{<,f}(t_1,t_2)\\
=\pm  {1}/{\hbar}\sum_{k=1}^{N}\bigg[ \left\langle \int_{t_0}^{t_1}dt'\xi_k(t')\;\hat{\rho}_{uv}^{f}(t_2)\hat{\rho}_{lj}^{f}(t_1)\hat{\rho}_{kk}^{f}(t')\right\rangle\\
+\left\langle \int_{t_1}^{t_2}dt'\xi_k(t')\;\hat{\rho}_{uv}^{f}(t_2)\hat{\rho}_{kk}^{f}(t')\hat{\rho}_{lj}^{f}(t_1)\right\rangle\\
+\left\langle \int_{t_2}^{t_0}dt'\xi_k(t')\;\hat{\rho}_{kk}^{f}(t')\hat{\rho}_{uv}^{f}(t_2)\hat{\rho}_{lj}^{f}(t_1)\right\rangle.
\end{multline}

After some simple calculations, we obtain
\begin{multline}
\delta g_{lj,vu}^{<,f}(t_1,t_2)=-\frac{i}{\hbar}\,g_{lj,uv}^{<,f}(t_1,t_2)\bigg[\int_{t_0}^{t_1}dt' \xi_j(t')\\
+\int_{t_1}^{t_2}dt' \xi_v(t')+\int_{t_2}^{t_0}dt' \xi_u(t')\bigg].
\end{multline}

\section{Calculation of transition rates}\label{app:fro}
Using Eqs.~(\ref{frozen}) and substituting in the first term of Eq. (\ref{eq:lam}), for $\kappa = 0$ we have for the imaginary part
\begin{multline}
i\int dt_1 \text{Im}\Big[g_{mj,vu}^{ r,f}(t,t_1)\Big]\Sigma_\alpha^{< (0)}(t_1,t)\\
=\pm \frac{\Gamma_\alpha(\Delta\epsilon_{uv})}{2}n_{\alpha}(\epsilon_{uv})\left(\left\langle \hat{\rho}^f_{mv}\right\rangle_t\delta_{ju}\pm \left\langle\hat{\rho}^f_{uj}\right\rangle_t\delta_{mv}\right),
\label{imag}
\end{multline}
which are referred to as {\em dissipation-type} terms\cite{limdblad}.
 Similarly, there are also terms of the type,
\begin{multline}
\int dt_1 \text{Re}\Big[g_{mj,vu}^{r,f}(t,t_1)\Big]\Sigma_\alpha^{< (0)}(t_1,t) \\
=-i\left\langle\left( \hat{\rho}^f_{mv}\right\rangle_t\delta_{ju}\pm \left\langle\hat{\rho}^f_{uj}\right\rangle_t\delta_{mv}\right)
\mathcal{P}\int\frac{d\epsilon}{2\pi}\frac{n_\alpha(\epsilon)\Gamma_\alpha(\epsilon)}{\epsilon-\epsilon_{uv}},
\label{real}
\end{multline}
which lead to {\em level renormalization}. For some specific spectral functions, the above integral can be calculated explicitly\cite{cuetara}.
Moreover, substituting Eq.~(\ref{frozen}) in the first term of Eq.~(\ref{eq:lam}), for $\kappa=1$ the imaginary part is given by
\begin{multline}
i\int dt_1 \text{Im}\Big[g_{mj,vu}^{ r,f}(t,t_1)\Big]\Sigma_\alpha^{< (1)}(t_1,t)\\
=\pm\epsilon_{uv} \frac{\Gamma_\alpha(\epsilon_{uv})}{2}n_{\alpha}(\epsilon_{uv})\left(\left\langle \hat{\rho}^f_{mv}\right\rangle_t\delta_{ju}\pm \left\langle\hat{\rho}^f_{uj}\right\rangle_t\delta_{mv}\right).
\label{imagen}
\end{multline}
 Similarly, the real part becomes
\begin{multline}
\int dt_1 \text{Re}\Big[g_{mj,vu}^{r,f}(t,t_1)\Big]\Sigma_\alpha^{< (1)}(t_1,t) \\
=-i\left\langle\left( \hat{\rho}^f_{mv}\right\rangle_t\delta_{ju}\pm \left\langle\hat{\rho}^f_{uj}\right\rangle_t\delta_{mv}\right)
\mathcal{P}\int\frac{d\epsilon}{2\pi}\frac{\epsilon\, n_\alpha(\epsilon)\Gamma_\alpha(\epsilon)}{\epsilon-\epsilon_{uv}}.
\label{realen}
\end{multline}
All other terms in Eq.~(\ref{eq:lam}) can be similarly evaluated. Substituting the above results in Eq.~(\ref{eq:lam}) neglecting the effect of lamb shift, for $\kappa=0$ we obtain
   \begin{multline}\label{lamfr}
\Lambda_{mj}^{\alpha (0)}(t)= \hbar\sum_u \Big[ \frac{\overline{\lambda}_{\alpha,ju}(t)}{2}\gamma_{\alpha,ujm}(t)\;{\rho}_{mu} \\
-\frac {\overline{\lambda}_{\alpha,u m}(t)}{2}\tilde{\gamma}_{\alpha,jmu}(t)\;{\rho}_{uj}\Big],
\end{multline}
and
\begin{multline}\label{overlamfr}
\overline{\Lambda}_{jm}^{\alpha (0)}(t)= \hbar\sum_u \Big[\frac{\lambda_{\alpha,ju}(t)}{2} \tilde{\bar{\gamma}}_{\alpha,mju}(t)\; {\rho}_{mu} \\
 - \frac{\lambda_{\alpha,um}(t)}{2}\bar{\gamma}_{\alpha,umj}(t)\;{\rho}_{uj} 
\Big].
\end{multline}
where we have introduced
\begin{eqnarray}
\gamma_{\alpha,ujm}(t)&=&\gamma_\alpha^f(\epsilon_{ju})+\delta\gamma_{\alpha,ujm}(t),\nonumber \\
\tilde{\gamma}_{\alpha,jmu}(t)&=&\tilde{\gamma}_\alpha^f(\epsilon_{um})+\delta\tilde{\gamma}_{\alpha,jmu}(t),\nonumber \\
\bar{\gamma}_{\alpha,umj}(t)&=&{\gamma}_\alpha^f(\epsilon_{mu})+\delta\bar{\gamma}_{\alpha,umj}(t),\nonumber \\
\tilde{\bar{\gamma}}_{\alpha,mju}(t)&=&\tilde{\gamma}_{\alpha}^f(\epsilon_{uj})+\delta\tilde{\bar{\gamma}}_{\alpha,mju}(t).
\label{eq:diff_gam}
\end{eqnarray}
The first terms on the right hand side of Eqs.~(\ref{eq:diff_gam}) are the frozen contributions originating from Eq.~(\ref{frozen}) and expressed as $\gamma_\alpha^f(\epsilon)=\hbar^{-1}n_\alpha(\epsilon)\Gamma^{(0)}_\alpha(\epsilon)$ and $\tilde{\gamma}_\alpha^f(\epsilon)=\hbar^{-1}(1\mp n_\alpha(\epsilon))\Gamma^{(0)}_\alpha(\epsilon)$. On the other hand, the second terms on the right hand side of Eqs.~(\ref{eq:diff_gam}) are due to the adiabatic correction given by Eq.~(\ref{dg}). The contribution due to level renormalization have been neglected.

\section{Evaluation of the mean values for the OTOCs}
\label{app:mean_otoc}
\begin{figure}[!htb]
	\centering
	\includegraphics[width=0.8\columnwidth]{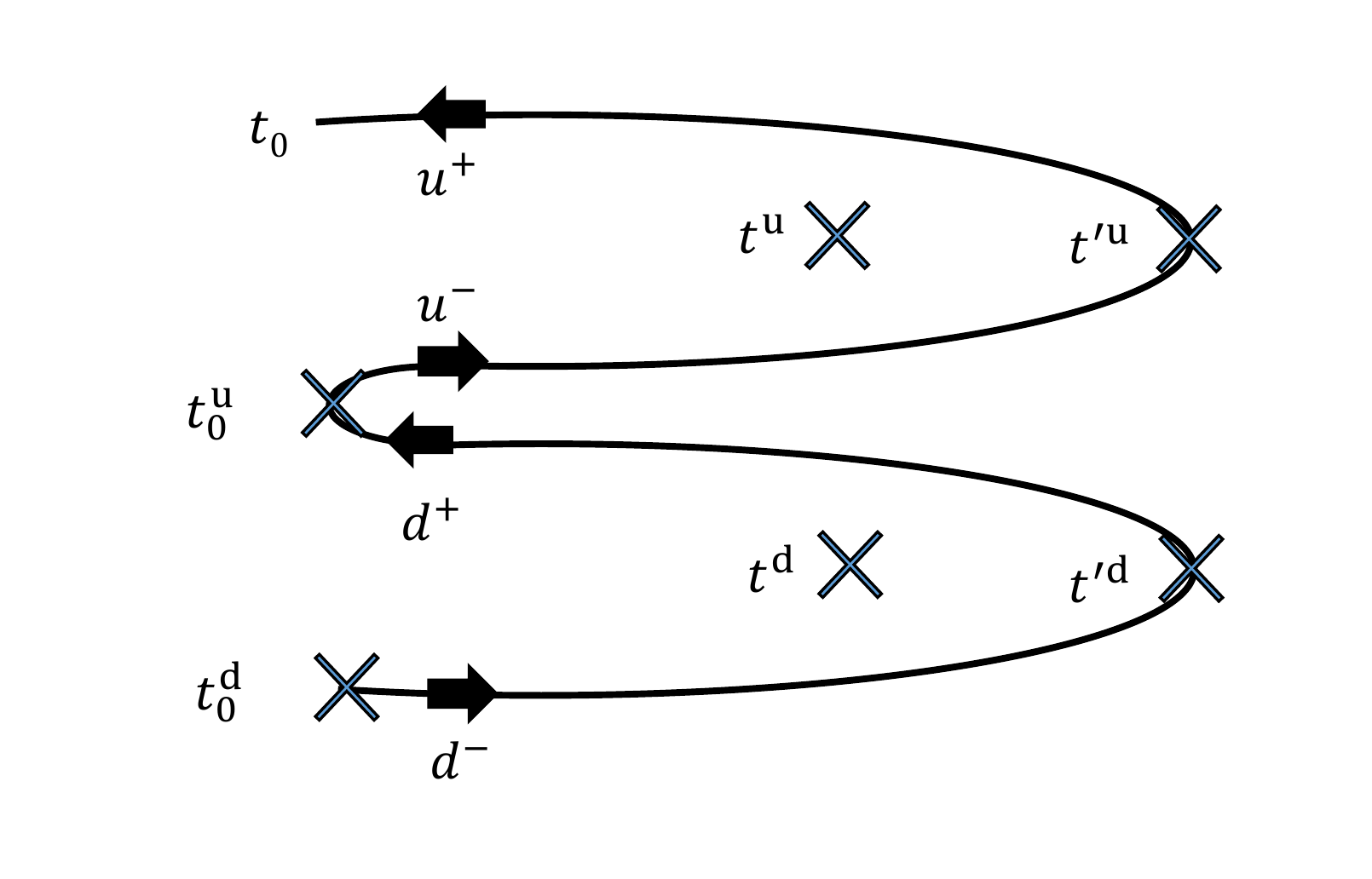}
	\caption{The augmented Keldysh contour $K$ for the OTOC.}
	\label{fig:aug_contour}
\end{figure}
\begin{figure}[b]
	\centering
	\includegraphics[width=0.8\columnwidth]{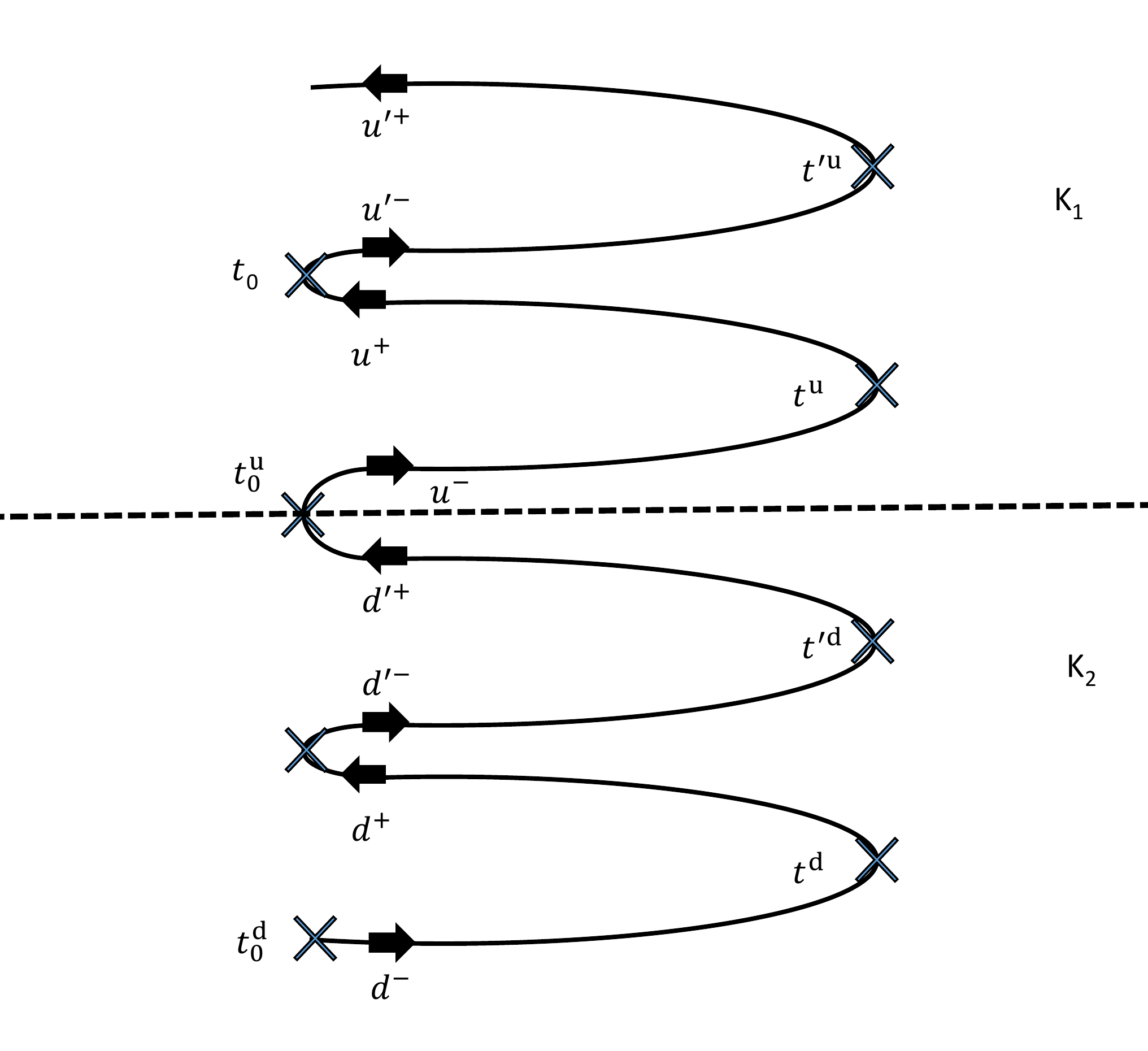}
	\caption{The deformed augmented Keldysh contour such that $t$ and $t'$ lie in different Keldysh contours.}
	\label{fig:contour_ana}
\end{figure}

For simplicity, we consider the bosonic case such that the bath and the system degrees of freedom commute.
The mean value associated with the lesser Green's function in the interaction picture can be written as
\begin{multline}
i\left\langle b_{k\alpha}^{\dagger{\cal H}} (t')\hat{K}_{lj,hf}(t)\right\rangle
\approx \int_K dt_1 \left\langle {T}_K\left[\tilde{H}_{\rm C}\,b_{k\alpha}^\dagger (t') \hat{K}_{lj,hf}(t)\right]\right\rangle
\\= \sum_{uv}V_{k\alpha}\bar{\lambda}_{\alpha,uv}(t)
 \int_K dt_1 \Big\langle {T}_K \Big[\hat{\rho}_{uv}(t_1)b_{k\alpha}(t_1)b_{k\alpha}^\dagger (t')\\
 \hat{\rho}_{lj}(t) {\cal O}_B(t_0)\hat{\rho}_{hf}(t){\cal O}_D(t_0)\Big]\Big\rangle,
\end{multline}
where $t$ lies in the arm where the contour goes from $-\infty$ to $\infty$ and $t^\prime$ in the arm which goes from $\infty$ to $-\infty$. The calculation of mean values cannot be done in the traditional Keldysh fashion as the out-of-time-order correlators (OTOC) have an abnormal time ordering. Instead we proceed along the line of argument of Ref.~\onlinecite{ioffe}. We will consider an augmented Keldysh contour as shown in Fig.~\ref{fig:aug_contour}. In terms of the augmented contour, the lesser Green's function can be expressed as
\begin{multline}
i\left\langle b_{k\alpha}^{\dagger{\cal H}} (t')\hat{K}_{lj,hf}^{\cal H}(t)\right\rangle\\
= \sum_{u,v}V_{k\alpha}\bar{\lambda}_{\alpha,uv}(t) \int_K dt_1 \Big\langle {T}_K \Big[\hat{\rho}_{uv}(t_1)b_{k\alpha}(t_1)b_{k\alpha}^\dagger ({t}^{\prime })\\
\hat{\rho}_{lj}(t^u) {\cal O}_B(t_0^u)\hat{\rho}_{hf}(t^d){\cal O}_D(t_0^d)\Big]\Big\rangle
\end{multline}
In the next step, we deform the contour as shown in Fig.~\ref{fig:contour_ana}.

The integration over the Keldysh contour can be broken down into 6 different parts ($d^-,~d^+,~u^-,~u^+$ in $K_1$ and forward and backward going branch in $K_2$) depending on where $t_1$ is pinned. The integral over the Keldysh contour can be expressed as
\begin{multline}
\int_K dt_1 = \int_{t_0^d}^{t^d}dt_1^{d^-}+\int_{t^d}^{t_0}dt_1^{d^+}+\int_{t_0}^{t'^{d}} dt_1^{d'^{-}}+\int_{t'^{d}}^{t_0^u} dt_1^{d'^{+}} \\
+\int_{t_0^u}^{t^u}dt_1^{u^-}+\int_{t^u}^{t_0}dt_1^{u^+}
+\int_{t_0}^{t^{\prime u }}dt_1^{u^{\prime -}}+\int_{t^{\prime  u}}^{t_0}dt_1^{u^{\prime +}}.
\end{multline}
We can stretch $t_0^{d/u}$ and $t_0$ to $-\infty$, such that 
\begin{widetext}
\begin{align}
i\left\langle b_{k\alpha}^{\dagger{\cal H}} (t')\hat{K}_{lj,hf}^{\cal H}(t)\right\rangle=& \sum_{u,v}V_{k\alpha}\bar{\lambda}_{\alpha,uv}(t) \int_{-\infty}^\infty dt_1\Bigg[\theta(t-t_1)\left\langle \hat{\rho}_{lj}(t){\cal O}_B(t_0)\left[\hat{\rho}_{hf}(t),\hat{\rho}_{uv}(t_1)\right]{\cal O}_D(t_0)\right\rangle\left\langle b_{k\alpha}^\dagger(t')b_{k\alpha}(t_1)\right\rangle
\nonumber \\
&+\theta(t'-t_1)\left\langle  \hat{\rho}_{lj}(t){\cal O}_B(t_0)\hat{\rho}_{uv}(t_1)\hat{\rho}_{hf}(t){\cal O}_D(t_0)\right\rangle\left[\left\langle b_{k\alpha}^\dagger(t')b_{k\alpha}(t_1)\right\rangle-\left\langle b_{k\alpha}(t_1)b_{k\alpha}^\dagger(t')\right\rangle\right]\nonumber \\
&+\theta(t-t_1)\left\langle  \left[\hat{\rho}_{lj}(t),\hat{\rho}_{uv}(t_1)\right]{\cal O}_B(t_0)\hat{\rho}_{hf}(t){\cal O}_D(t_0)\right\rangle\left\langle b_{k\alpha}^\dagger(t')b_{k\alpha}(t_1)\right\rangle\nonumber \\
&+\theta(t'-t_1)\left\langle  \hat{\rho}_{uv}(t_1)\hat{\rho}_{lj}(t){\cal O}_B(t_0)\hat{\rho}_{hf}(t){\cal O}_D(t_0)\right\rangle\left[\left\langle b_{k\alpha}^\dagger(t')b_{k\alpha}(t_1)\right\rangle-\left\langle b_{k\alpha}(t_1)b_{k\alpha}^\dagger(t')\right\rangle\right]\Bigg].
\label{eq:exp_kel}
\end{align}
\end{widetext}
In terms of Green's functions, one can write
\begin{multline}
G_{ljhf,k\alpha}^<(t,t') \simeq \int_{-\infty}^\infty dt_1 V_{k\alpha} \sum_{u,v}\overline{\lambda}_{\alpha,uv}(t)\\
\Big[g_{ljhf,uv}^r(t,t_1)g_{k\alpha}^<(t_1,t') +g_{ljhf,uv}^<(t,t_1)g_{k\alpha}^a(t_1,t')\Big],
\end{multline}
where
\begin{multline}
g^r_{ljhf,vu}(t,t')\\
=-i\theta(t-t')\Big\langle\hat{\rho}_{lj}(t){\cal O}_B(t_0)
\Big[\hat{\rho}_{hf}(t),\hat{\rho}_{uv}(t')\Big]_-\\{\cal O}_D(t_0)+\Big[\hat{\rho}_{lj}(t),\rho_{uv}(t')\Big]_-{\cal O}_B(t_0)\hat{\rho}_{hf}(t){\cal O}_D(t_0)\Big\rangle,\\
\end{multline}
and 
\begin{multline}
g^<_{ljhf,vu}(t,t')=- i\left\langle\hat{\rho}_{uv}(t') \hat{\rho}_{lj}(t){\cal O}_B(t_0) \hat{\rho}_{hf}(t){\cal O}_D(t_0)\right\rangle\\
- i\left\langle \hat{\rho}_{lj}(t){\cal O}_B(t_0)\hat{\rho}_{uv}(t') \hat{\rho}_{hf}(t){\cal O}_D(t_0)\right\rangle.
\end{multline}
In the case with fermionic baths, the commutator changes to anti-commutator and the lesser Green's function changes by a sign.
\section{Frozen and Adiabatic components of Lesser Green's function for the case of OTOC}
The lesser Green's function for the case of OTOC can be separated into frozen and adiabatic components (similar to the case of density matrix).
We now introduce the interaction representation with respect to $\tilde{{\cal H}}^f_{\rm S}$ and consider the  Green's function
\begin{align}
g_{ljhf,vu}^<(t_1,t_2)=g_{ljhf,uv}^{<,f}(t_1,t_2)+\delta g_{ljhf,uv}^{<,f}(t_1,t_2),
\end{align}
where the frozen contribution is
\begin{multline}
g_{ljhf,vu}^{<,f}(t_1,t_2)=\pm i \left\langle \hat{\rho}_{uv}^f (t_2^-)\hat{K}_{lj,hf}^f(t_1^+)\right\rangle \\
\pm i \left\langle \hat{\rho}_{lj}^f(t_1^+){\cal O}_B(t_0)\hat{\rho}_{uv}^f(t_2^-)\hat{\rho}_{hf}^f(t_1^+){\cal O}_D(t_0)\right\rangle
\end{multline}
and the adiabatic contribution is
\begin{multline}
\delta g_{ljhf,vu}^{<,f}(t_1,t_2)\\
=\pm \frac{1}{\hbar} \Big\langle T_{k} \int_K dt' \delta \tilde{H}_S^f(t')\hat{\rho}_{uv}^f (t_2^-)\hat{K}_{lj,hf}^f(t_1^+)\Big\rangle \pm \frac{1}{\hbar} \Big\langle T_{k} \int_K dt' \\
\delta \tilde{H}_S^f(t')\hat{\rho}_{lj}^f(t_1^+){\cal O}_B(t_0)\hat{\rho}_{uv}^f(t_2^-)\hat{\rho}_{hf}^f(t_1^+){\cal O}_D(t_0)\Big\rangle
\end{multline}
After some calculation using the Keldysh contour in Fig.~\ref{fig:aug_contour}, we obtain
\begin{multline}
\delta g_{ljhf,vu}^{<,f}(t_1,t_2)=\pm \frac{1}{\hbar}\int_{-\infty}^\infty dt' \bigg(\left\langle \hat{K}_{uj,hf}^f(t_1)\right\rangle \delta_{vl}\times\\
\Big[\theta(t_1-t')\xi_{fj,hv}(t')+\theta(t_2-t')\xi_{v,u}(t')\Big]+\left\langle \hat{K}_{lj,uf}^f(t_1)\right\rangle\\
\delta_{vh}\Big[\theta(t_1-t')\xi_{fj,lv}(t')+\theta(t_2-t')\xi_{v,u}(t')\Big]\bigg)e^{i\epsilon_{vu}(t_1-t_2)}.
\end{multline}

\section{Calculation of frozen transition rates for the OTOC master equation}
\label{app:tran_otoc}
We have the following relations for the bath Green's function
\begin{align}
g^r_{k\alpha}(t,t_1)&=-i\theta (t-t_1)e^{-i\epsilon_{k\alpha}(t-t_1)},\nonumber\\
g^a_{k\alpha}(t,t_1)&=i\theta (t_1-t)e^{-i\epsilon_{k\alpha}(t-t_1)},\nonumber \\
g^<_{k\alpha}(t,t_1)&=\pm i n_\alpha(\epsilon_{k\alpha})e^{-i\epsilon_{k\alpha}(t-t_1)}.
\end{align}
Similarly for the system Green's functions
\begin{multline}
g_{ljl'j',vu}^{<, f}(t,t_1)= 
 \pm i\, \Big[\delta_{lv} \left\langle K_{uj,l'j'}^f(t)\right\rangle \\
 +\delta_{l'v} \left\langle K_{lj,uj'}^f(t)\right\rangle\Big]e^{i\epsilon_{vu}(t-t_1)},
\end{multline}
\begin{multline}
g_{uv,jlj'l'}^{<, f}(t,t_1)= 
 \pm i\, \Big[\delta_{j'u}\left\langle K_{lj,l'v}^f(t)\right\rangle \\
 +\delta_{ju} \left\langle K_{lv,l'j'}^f(t)\right\rangle\Big]e^{i\epsilon_{vu}(t-t_1)},
\end{multline}
the retarded Green's function
\begin{multline}
g^{r,f}_{ljl'j',vu}(t,t_1)=-i\theta(t-t_1)e^{i\epsilon_{vu}(t-t_1)}\Big(\delta_{uj}\left\langle K_{lv,l'j'}^f(t)\right\rangle \\
\pm \delta_{vl}\left\langle K_{uj,l'j'}^f(t)\right\rangle 
+\delta_{j'u}\left\langle K_{lj,l'v}^f(t)\right\rangle\pm \delta_{vl'}\left\langle K_{lj,uj'}^f(t)\right\rangle\Big),
\end{multline}
and the advanced Green's function
\begin{multline}
g_{uv,jlj'l'}^{a,f}(t_1,t)=i\theta(t-t_1)e^{i\epsilon_{vu}(t-t_1)}\Big(\delta_{vl}\left\langle K_{uj,l'j'}^f(t)\right\rangle\\
\pm \delta_{uj}\left\langle K_{lv,l'j'}^f(t)\right\rangle+\delta_{vl'}\left\langle K_{lj,uj'}^f(t)\right\rangle\pm \delta_{j'u}\left\langle K_{lj,l'v}^f(t)\right\rangle\Big).
\end{multline}
Now we can calculate individual expressions in Eq.~24 such as
\begin{multline}
\int dt_1 g_{ljl'j',vu}^{r,f}(t,t_1)g_{k\alpha}^<(t_1,t)\\
=\pm i \Big(\delta_{uj}\left\langle K_{lv,l'j'}^f(t)\right\rangle 
\pm \delta_{vl}
\left\langle K_{uj,l'j'}^f(t)\right\rangle 
+\delta_{j'u}\left\langle K_{lj,l'v}^f(t)\right\rangle\\
\pm \delta_{vl'}\left\langle K_{lj,uj'}^f(t)\right\rangle\Big) n_{\alpha}(\epsilon_{k\alpha})\left[\frac{1}{\epsilon_{k\alpha}-\epsilon_{uv}+i\eta}\right].
\end{multline}

Similarly,
\begin{multline}
\int dt_1 g_{ljl'j',vu}^{<,f}(t,t_1)g_{k\alpha}^a(t_1,t)=\mp i \Big[\delta_{vl}\left\langle K_{uj,l'j'}^f(t)\right\rangle \\ +
\delta_{vl'}\left\langle K_{lj,uj'}^f(t)\right\rangle\Big]\left[\frac{1}{\epsilon_{k\alpha}-\epsilon_{uv}+i\eta}\right],
\end{multline}
\begin{multline}
\int_{-\infty}^\infty dt_1 g_{k\alpha}^r(t,t_1)g_{uv,jlj'l'}^{<,f}(t_1,t)=\mp i\Big[\delta_{j'u}\left\langle K_{lj,l'v}^f(t)\right\rangle\\
+\delta_{ju}\left\langle K_{lv,l'j'}^f(t)\right\rangle\Big]\left[\frac{1}{\epsilon_{k\alpha}-\epsilon_{vu}+i\eta}\right],
\end{multline}
and
\begin{multline}
\int dt_1 g_{k\alpha}^<(t,t_1)g_{uv,jlj'l'}^{a,f}(t_1,t)\\
=\pm i \Big(\delta_{vl}\left\langle K_{uj,l'j'}^f(t)\right\rangle \pm \delta_{uj}\left\langle K_{lv,l'j'}^f(t)\right\rangle+\delta_{vl'}\left\langle K_{lj,uj'}^f(t)\right\rangle\\
\pm \delta_{j'u}\left\langle K_{lj,l'v}^f(t)\right\rangle\Big)n_{\alpha}(\epsilon_{k\alpha})\left[\frac{1}{\epsilon_{k\alpha}-\epsilon_{vu}+i\eta}\right].
\end{multline}
Using the relation
\begin{equation}
\frac{1}{\epsilon_{k\alpha}-\epsilon_{mn}\pm i\eta}={\mathcal P}\left\{\frac{1}{\epsilon_{k\alpha}-\epsilon_{mn}}\right\} \mp i\pi \delta (\epsilon_{k\alpha}-\epsilon_{mn}),  
\end{equation}
and neglecting the principal value (which gives rise of level renormalization effects), we obtain
\begin{multline}
\int dt_1 g_{ljl'j',vu}^{r,f}(t,t_1)\Sigma_{\alpha}^{<(0)}(t_1,t)=\pm \frac{\Gamma_\alpha^{(0)}(\epsilon_{uv})}{2}n_\alpha(\epsilon_{uv})\\
\Big(\delta_{uj}\left\langle K_{lv,l'j'}^f(t)\right\rangle 
\pm \delta_{vl}
\left\langle K_{uj,l'j'}^f(t)\right\rangle 
+\delta_{j'u}\left\langle K_{lj,l'v}^f(t)\right\rangle\\
\pm \delta_{vl'}\left\langle K_{lj,uj'}^f(t)\right\rangle\Big),
\label{eq:lamotoc1}
\end{multline}
where $\Sigma_{k\alpha}^{(0)}(t_1,t)=\sum_k |V_{k\alpha}|^2g_{k\alpha}(t_1,t)$ for all the bath Green's functions. Similarly,
\begin{multline}
\int dt_1 g_{ljl'j',vu}^{<,f}(t,t_1)\Sigma_\alpha^{a(0)}(t_1,t)\\
=\mp \frac{\Gamma_\alpha^{(0)}(\epsilon_{uv})}{2}\Big[\delta_{vl}\left\langle K_{uj,l'j'}^f(t)\right\rangle  +
\delta_{vl'}\left\langle K_{lj,uj'}^f(t)\right\rangle\Big],
\label{eq:lamotoc2}
\end{multline}
\begin{multline}
\int dt_1 \Sigma_\alpha^{r(0)}(t,t_1)g_{uv,jlj'l'}^{<,f}(t_1,t)\\
=\mp \frac{\Gamma_\alpha^{(0)}(\epsilon_{vu})}{2}\Big[\delta_{j'u}\left\langle K_{lj,l'v}^f(t)\right\rangle
+\delta_{ju}\left\langle K_{lv,l'j'}^f(t)\right\rangle\Big],
\label{eq:lamotoc3}
\end{multline}
and
\begin{multline}
\int dt_1 \Sigma_\alpha^{<(0)}(t,t_1)g_{uv,jlj'l'}^{a,f}(t_1,t)=\pm \frac{\Gamma_\alpha^{(0)}(\epsilon_{vu})}{2}n_{\alpha}(\epsilon_{vu})\\
\Big(\delta_{vl}\left\langle K_{uj,l'j'}^f(t)\right\rangle \pm \delta_{uj}\left\langle K_{lv,l'j'}^f(t)\right\rangle+\delta_{vl'}\left\langle K_{lj,uj'}^f(t)\right\rangle\\
\pm \delta_{j'u}\left\langle K_{lj,l'v}^f(t)\right\rangle\Big).
\label{eq:lamotoc4}
\end{multline}
Substituting Eqs.~(\ref{eq:lamotoc1}),~(\ref{eq:lamotoc2}),~(\ref{eq:lamotoc3}) and~(\ref{eq:lamotoc4}) in Eqs.~(\ref{eq:mixed1}) and (\ref{eq:mixed2}), in terms of transition rates we obtain
\begin{widetext}
\begin{multline}
{\Lambda}^{\alpha, \rm OTOC}_{ljl^{\prime} j^{\prime}}(t,t)=\frac{\hbar}{2}\sum_u\bigg[\overline{\lambda}_{\alpha,ju}(t)\gamma_\alpha(\epsilon_{ju})\left\langle K_{lu,l'j'}^f(t)\right\rangle +\overline{\lambda}_{\alpha,j'u}(t)\gamma_\alpha(\epsilon_{j'u})\left\langle K_{lj,l'u}^f(t)\right\rangle\\
-\overline{\lambda}_{\alpha,ul}(t)\tilde{\gamma}_\alpha(\epsilon_{ul})\left\langle K_{uj,l'j'}^f(t)\right\rangle -\overline{\lambda}_{\alpha,ul'}(t)\tilde{\gamma}_{\alpha}(\epsilon_{ul'})\left\langle K_{lj,uj'}^f(t)\right\rangle\bigg],
\end{multline}
\begin{multline}
\overline{\Lambda}^{\alpha, \rm OTOC}_{jl j^{\prime}l^\prime}(t,t)=-\frac{\hbar}{2}\sum_u\bigg[\lambda_{\alpha,ul}(t)\gamma_\alpha(\epsilon_{lu})\left\langle K_{uj,l'j'}^f(t)\right \rangle +\lambda_{\alpha,ul'}(t)\gamma_{\alpha}(\epsilon_{l'u})\left\langle K_{lj,uj'}^f(t)\right\rangle\\
-\lambda_{\alpha,j'u}(t)\tilde{\gamma}_\alpha(\epsilon_{uj'})\left\langle K_{lj,l'u}^f(t)\right\rangle - \lambda_{\alpha,ju}(t)\tilde{\gamma}_{\alpha}(\epsilon_{uj})\left\langle K_{lu,l'j'}^f(t)\right\rangle\bigg].
\end{multline}
Finally, we have
\begin{multline}
\frac{\partial  K_{lj,l'j'}(t)}{\partial t}=\frac{i}{\hbar}\left[\epsilon_l(t)-\epsilon_j(t)+\epsilon_{l'}(t)-\epsilon_{j'}(t)\right] K_{lj,l'j'}(t)+\frac{1}{\hbar}\sum_m\Big[\lambda_{\alpha,ml}(t)\Lambda^{\alpha,{\rm OTOC}}_{mjl'j'}(t)-\lambda_{\alpha,jm}(t)\Lambda^{\alpha,{\rm OTOC}}_{lml'j'}(t)\\
+\lambda_{\alpha,ml'}(t)\Lambda^{\alpha,{\rm OTOC}}_{ljmj'}(t)-\lambda_{\alpha,j'm}(t)\Lambda^{\alpha,{\rm OTOC}}_{ljl'm}(t)
+\overline{\lambda}_{\alpha,ml}(t)\overline{\Lambda}^{\alpha,{\rm OTOC}}_{jmj'l'}(t)-\overline{\lambda}_{\alpha,jm}(t)\overline{\Lambda}^{\alpha,{\rm OTOC}}_{mlj'l'}(t)\\
+\overline{\lambda}_{\alpha,ml'}(t)\overline{\Lambda}^{\alpha,{\rm OTOC}}_{jlj'm}(t)-\overline{\lambda}_{\alpha,j'm}(t)\overline{\Lambda}^{\alpha,{\rm OTOC}}_{jlml'}(t)\Big]
\end{multline}

\end{widetext}

\section{Adiabatic  change of the basis of eigenstates}\label{apgau}
Introducing the notation $\partial_{\bf X}$ instead of $\partial/\partial_{\bf X}$, we have,
\begin{equation}
\partial_{\bf X} \left(|l\rangle\langle j| \right)=|\partial_{\bf X} l \rangle \langle j|+|l\rangle \langle \partial_{\bf X} j|
\end{equation}
Using Eq.~D3 of Ref.~\onlinecite{adiageo}, we obtain
\begin{equation}
\langle l'| \partial_{\bf X} l\rangle =\frac{\langle l'| \partial_ {\bf X} {\cal H}_S |l\rangle}{\varepsilon_l-\varepsilon_{l'}}, ~~~l\neq l'.
\end{equation}
Operating on both sides by $\sum_{l'}|l'\rangle$, and taking only the contribution originating in the gauge term, we obtain
\begin{equation}
\partial_{\bf X}|l\rangle = \sum_{l'}A_{l',l}|l'\rangle.
\end{equation}
Similarly, using the other equation in Eq.~D3 of Ref.~\onlinecite{adiageo}, we obtain
\begin{equation}
\langle\partial_{\bf X} j|l'\rangle=\frac{\langle j| \partial_ {\bf X} {\cal H}_S |l'\rangle}{\varepsilon_j-\varepsilon_{l'}}, ~~~l\neq l'.
\end{equation}
Operating on both sides by $\sum_{l'}\langle l'|$, we obtain
\begin{equation}
\langle\partial_{\bf X} j|=-\sum_{l'}A_{j,l'}\langle l'|
\end{equation}

\bibliography{mas_kel_bib}
\end{document}